\documentclass[conference]{IEEEtran}
\IEEEoverridecommandlockouts
\usepackage{cite}
\usepackage{amsmath,amssymb,amsfonts}
\usepackage{algorithmic}
\usepackage{dblfloatfix}
\usepackage{graphicx}
\usepackage{placeins}
\usepackage{textcomp}
\usepackage{color,soul}
\usepackage{dblfloatfix}
\usepackage[table,xcdraw]{xcolor}
\usepackage[listings]{tcolorbox}
\usepackage{xcolor}
\def\BibTeX{{\rm B\kern-.05em{\sc i\kern-.025em b}\kern-.08em
    T\kern-.1667em\lower.7ex\hbox{E}\kern-.125emX}}
\usepackage[table]{xcolor}
\usepackage{enumitem}
\usepackage{hyperref}
\usepackage{xcolor}
\usepackage{fontawesome}
\usepackage{multirow}
\usepackage{subcaption}
\usepackage{balance}

\usepackage{pgfplots}
\usepackage{pgfplotstable}
\pgfplotsset{compat=1.7}
\usepackage{tikz}
\hypersetup{
  colorlinks   = True, 
  urlcolor     = blue, 
  linkcolor    = blue, 
  citecolor    = blue 
}

\title{Engineering Software Systems for Quantum Computing as a Service: A Mapping Study}

\author{
\IEEEauthorblockN{Aakash Ahmad\IEEEauthorrefmark{1}, Muhammad Waseem\IEEEauthorrefmark{2}, Peng Liang\IEEEauthorrefmark{3}, Mahdi Fehmideh\IEEEauthorrefmark{4}}
\IEEEauthorblockN{Arif Ali Khan\IEEEauthorrefmark{5}, David Georg Reichelt\IEEEauthorrefmark{1}, Tommi Mikkonen\IEEEauthorrefmark{2}}
\IEEEauthorblockA{\IEEEauthorrefmark{1} School of Computing and Communications, Lancaster University Leipzig, Leipzig, Germany}
\IEEEauthorblockA{\IEEEauthorrefmark{2} Faculty of Information Technology, University of Jyväskylä, Jyväskylä, Finland}
\IEEEauthorblockA{\IEEEauthorrefmark{3} School of Computer Science, Wuhan University, Wuhan, China}
\IEEEauthorblockA{\IEEEauthorrefmark{4} School of Business at University of Southern Queensland, Queensland, Australia}
\IEEEauthorblockA{\IEEEauthorrefmark{5} M3S Empirical Software Engineering Research Unit, University of Oulu, Oulu, Finland}
\IEEEauthorblockA{a.ahmad13@lancaster.ac.uk, mwaseem@jyu.fi, liangp@whu.edu.cn, mahdi.fahmideh@usq.edu.au}
\IEEEauthorblockA{arif.khan@oulu.fi, d.g.reichelt@lancaster.ac.uk, tommi.j.mikkonen@jyu.fi}
}

\begin{document}
\maketitle

\begin{abstract}
Quantum systems have started to emerge as a disruptive technology and enabling platforms – exploiting the principles of quantum mechanics – to achieve quantum supremacy in computing. Academic research, industrial projects (e.g., Amazon Braket), and consortiums like `Quantum Flagship’ are striving to develop practically capable and commercially viable quantum computing (QC) systems and technologies. Quantum Computing as a Service (QCaaS) is viewed as a solution attuned to the philosophy of service-orientation that can offer QC resources and platforms, as utility computing, to individuals and organisations who do not own quantum computers. To understand the quantum service development life cycle and pinpoint emerging trends, we used evidence-based software engineering approach to conduct a systematic mapping study (SMS) of research that enables or enhances QCaaS. The SMS process retrieved a total of 55 studies, and based on their qualitative assessment we selected 9 of them to investigate (i) the functional aspects, design models, patterns, programming languages, deployment platforms, and (ii) trends of emerging research on QCaaS. The results indicate three modelling notations and a catalogue of five design patterns to architect QCaaS, whereas Python (native code or frameworks) and Amazon Braket are the predominant solutions to implement and deploy QCaaS solutions. 
From the quantum software engineering (QSE) perspective, this SMS provides empirically grounded findings that could help derive processes, patterns, and reference architectures to engineer software services for QC.
\end{abstract}

\begin{IEEEkeywords}
 Quantum Software Engineering, Quantum Service Computing, Systematic Mapping Study, Software Services.
\end{IEEEkeywords}

\section{Introduction}
\label{sec:introduction}
Quantum computing (QC) has started to emerge as a disruptive technology and an enabling platform – exploiting the principles of quantum mechanics – relying on Quantum Bits (QuBits) that manipulate Quantum Gates (QuGates) to tackle computationally intensive tasks efficiently \cite{R1_ali2022software}. QC systems are in a phase of continuous evolution and despite being in a state of their infancy, such systems have started to computationally outperform their classical counterparts (i.e., digital computers) in applications such as quantum information processing, bio-inspired computing, and simulation of quantum mechanics \cite{R2_harrow2017quantum}. Academic research \cite{R1_ali2022software} \cite{R3_egger2020quantum} and industrial initiatives led by technology giants such as IBM, Google, and Microsoft \cite{R4_dyakonov2019will} are striving hard to achieve strategic advantages associated with quantum systems and software technologies in a so-called `race to quantum economy’. A recent report presented at the World Economic Forum titled \textit{State of Quantum Computing: Building a Quantum Economy} highlights that by the year 2022, public and private investments in quantum computing technologies totalled \$35.5 billion \cite{R5_WorldEconomicForum}. Despite the strategic capabilities that can be attained via quantum supremacy in computing; programming, operationalising, and maintaining a quantum computer is a complex and radically distinct engineering paradigm \cite{R1_ali2022software}. To augment the quantum hardware development, state-funded projects, and global consortiums are proactively funding initiatives such as the Quantum Flagship \cite{R14_riedel2019europe} and National Quantum Initiative \cite{R22_raymer2019us} to develop software ecosystems, networking technologies, and human expertise for the alleged quantum leap in computing \cite{R13_monroe2018quantum}.

\begin{figure}[]
 \centering
 \includegraphics[scale=0.80]{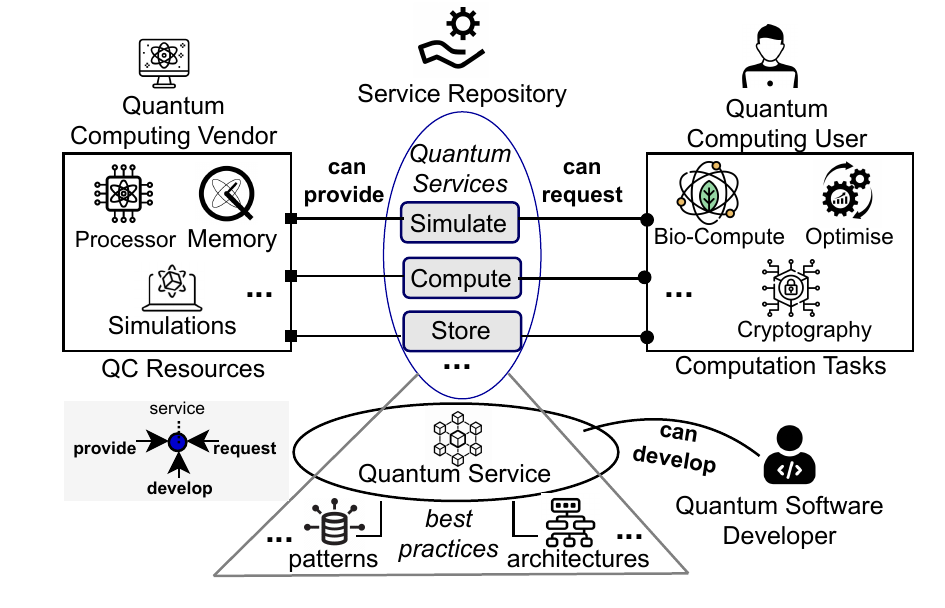} 
 	\caption{A Generic View of the QCaaS Model}
	\label{Fig-2:QCaaS}
\end{figure}

\textbf{Motivation: Pay-per-use QCaaS -} Service-oriented systems have proven to be useful in supporting utility computing model that relies on pay-per usability approach for individuals and organisations to utilize a multitude of computing services without the need to own or maintain them \cite{R6_wei2010service}. Since their early adoption, pay-per-use service-driven systems have now grown from data storage, video streaming, and entertainment, to resource-sharing applications that represent a multi-billion dollar industry in service economies \cite{R7_bouguettaya2017service}\cite{R8_Statista}. The ‘as-a-Service’ (aaS) model has provided the impetus to wide-scale adoption of service systems that can offer a plethora of computing services including but not limited to storage, computation, infrastructure, platform, and software to end-users \cite{R7_bouguettaya2017service}. The aaS model can enable users and developers who can exploit the QC platforms (e.g., processors, memory, simulators) offered by quantum vendors, such as Amazon, Goole, and IBM \cite{R9_moguel2022quantum}. Quantum Computing as a Service (QCaaS) as in Figure \ref{Fig-2:QCaaS} is a recent and quantum-specific genre of aaS model that is built on the philosophy of service-orientation of QC, i.e., pay-per-shot at QC resources instead of owning, programming, and/or maintaining quantum computers \cite{R10_garcia2021quantum}. Quantum vendors view pay-per-shot as an opportunistic business model to generate revenue streams from their QC infrastructures, where a shot is a single execution of a quantum algorithm on a quantum processing unit (QPU). 
The QCaaS can alleviate the need to own or maintain quantum computers and can help software and service developers  to rely on  existing knowledge and best practices to develop software services  that can be executed on QC platforms \cite{R11_leymann2020quantum}.  Quantum software services enable developers to wrap data and computations inside loosely coupled, fine-grained modules of source code to execute tasks such as prime factorisation, key encryption, or bio stimulations on QC platforms \cite{R10_garcia2021quantum}. There is a growing interest in the academic community and industrial vendors to research and develop solutions for enabling quantum service-orientation \cite{R9_moguel2022quantum}\cite{R10_garcia2021quantum}.

\textbf{Needs for the mapping study:} Systematic mapping studies (SMS) rely on evidence-based software engineering approach to systematically and reproducibly identify, analyse, synthesise, and document (i.e., map the trends of) existing research on the topic under investigation \cite{R16_petersen2008systematic}. The academic community aims to exploit QC platforms for empiricism in quantum research \cite{R9_moguel2022quantum}, while the vendors pursue a revenue stream as well as validation and testing of their under-development quantum platforms, offered as a service \cite{R10_garcia2021quantum}. Synergising research and development on quantum computing with service-orientation can allow researchers and developers to leverage existing design principles, patterns, architectural styles, and modelling languages of services computing to offer QC as a service \cite{R11_leymann2020quantum}. Moreover, systemizing the efforts to architect and implement QaaS requires discovering new patterns and developing innovative frameworks rooted in empirically grounded guidelines to research emerging challenges and develop futuristic solutions \cite{R12_valencia2022quantum}. We conducted this SMS based on two research questions to investigate (i) \textit{existing solutions in terms of designing, implementing, deploying, and operationalizing quantum software services} and  (ii) \textit{ emerging trends that can indicate futuristic research on QCaaS}. 
The results indicate that during quantum service-orientation, software modeling languages (e.g., UML) and patterns (e.g., service wrapping, API gateway) help in mapping functional requirements to service implementation (e.g., Python code) that can be deployed on QC platforms (e.g., Amazon Braket). Emerging trends indicate non-functional aspects, model-driven engineering (low-code development), empirically discovered tactics, human roles, and process-centric development of QCaaS.

\textbf{Contributions and implications} of this SMS are to: 

\begin{itemize}
    \item identify and document a collective impact of existing research (published academic evidence)  to investigate the extent to which classical and quantum-specific service-orientation can be applied to QCaaS.
    \item highlight emerging trends - identifying existing gaps - that reflect the dimensions of future research to develop emerging and next-generation of QCaaS solutions.      
\end{itemize}

Academic researchers can rely on mapping of existing research and vision for future work to research and develop QCaaS solutions in a broader context of QSE \cite{R1_ali2022software}\cite{R14_riedel2019europe}. Practitioners who can rely on academic references about patterns (reusability), modeling notations (representation), and implementation (prototyping) to develop QCaaS \cite{R9_moguel2022quantum}.    


\section{Research Context and Method}
\label{ResearchContextMethod}
We now contextualise service-orientation for QCs in Section \ref{sec:context} and present the research method in Section \ref{sec:method}. 

\subsection{Context: Service-Orientation for Quantum Computing}
\label{sec:context}
\subsubsection{Quantum Computing Systems}
\begin{figure}[]
 \centering
 \includegraphics[scale=0.45]{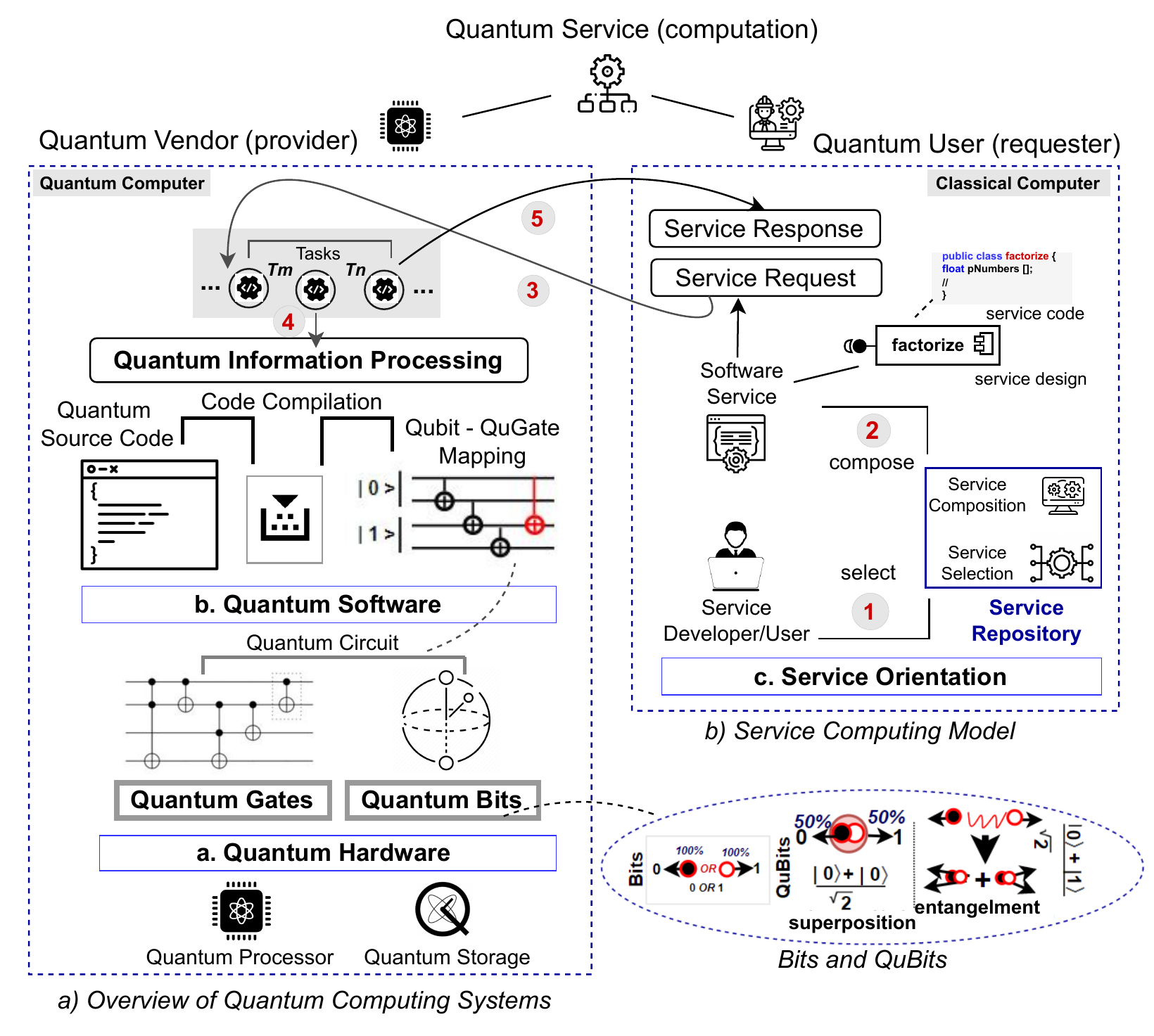} 
 	\caption{An Overview of the Quantum Service-Orientation}
	\label{Fig-1:Context}
\end{figure}

We briefly overview a QC system that comprises of quantum hardware and software elements as shown in Figure \ref{Fig-1:Context}a). Fundamental to achieving quantum computations are Quantum Bits (QuBits) that represent the basic unit of quantum information processing by manipulating Quantum Gates (QuGates) \cite{R2_harrow2017quantum} \cite{R13_monroe2018quantum}. 
Traditional Binary Digits (Bits) in classical systems (i.e., digital computers) are represented as [1, 0] where 1 represents the computation state as \textsc{On} and 0 represents the state as \textsc{Off} to manipulate binary gates in digital circuits. In comparison, a QuBit represents a two-state quantum computer expressed as $|0\rangle$ and $|1\rangle$. The state of a single Qubit can be expressed as  $|0\rangle = \begin{bmatrix} 1 \\ 0 \end{bmatrix}$ and $|1\rangle = \begin{bmatrix} 0 \\ 1 \end{bmatrix}$ and quantum superposition allows a QuBit to attain a liner combination of both states: 

\begin{equation}\label{EQ-1}
 |0\rangle  =  \left[ \begin{array}{c} 1 \\ 0 \end{array} \right] ~~~~~ + ~~~~~  |1\rangle  =  \left[ \begin{array}{c} 0 \\ 1 \end{array} \right]   
\end{equation}

Based on Figure \ref{Fig-1:Context}a), we distinguish between a Bit and QuBit such that a Bit can take a value as either \textsc{`Off:0’} or \textsc{`On:1’} with 100\% probability. In comparison, a QuBit can be in a state of $|0\rangle$ or $|1\rangle$ or in a superposition state with 50\% $|0\rangle$ and 50\%  $|1\rangle$. In addition, two QuBits can be entangled, and entangled QuBits are linked in a way that observing (i.e., measuring) one of the QuBits, can reveal the state of the other QuBit. Extended details about QuBits and QuGates to develop and operate the QC systems are reported in studies like \cite{R1_ali2022software} and \cite{R9_moguel2022quantum}. To utilise the quantum computing resources, such as quantum processor and memory, there is a need for control software that can program QuBits to manage QuGates of a QC system. Quantum software systems rely on quantum source code compilers that allow quantum algorithm designers and programmers to write, build, and execute software for quantum computers. For example, a programmer can use a quantum programming language, such as Q\# (by Microsoft) or Qiskit (by Google) and use a quantum compiler to enable programable quantum computations \cite{R4_dyakonov2019will} \cite{R15_de2022software}. Software systems that can manage and control quantum hardware find their applications in areas including but not limited to quantum cryptography, bio-inspired computing, and quantum information processing \cite{R1_ali2022software}. However, scarcity of quantum hardware, lack of quantum software professionals, and economics of owning or maintaining QC are some critical factors that impede commercially viable quantum computers \cite{R2_harrow2017quantum} \cite{R3_egger2020quantum}. Vendors who offer QCaaS platforms view quantum service-orientation as an opportunistic business model that offers QC resources to customers as utility computing \cite{R6_wei2010service} \cite{R9_moguel2022quantum}.

\subsubsection{Service-Oriented Computing}
Services computing follows the SOA style that allows service users to discover and utilise a multitude of available software services that encapsulate computing resources and applications offered by service vendors/providers \cite{R7_bouguettaya2017service}. Figure \ref{Fig-1:Context}b) shows SOA style-based quantum servicing where a QC user (i.e., service requester) can utilise the QC resources offered by quantum vendors (i.e., service provider) by means of quantum services. In most cases, QC systems of today are not capable of executing quantum algorithms wrapped with large amounts of data, inputs, and outputs \cite{R2_harrow2017quantum} \cite{R13_monroe2018quantum}. As shown in Figure \ref{Fig-1:Context}a), large volumes of data in quantum algorithms require more QuBits and complex QuGates that result in deep quantum circuiting and consequently increased errors referred to as noisy intermediate-scale quantum (NISQ) \cite{R4_dyakonov2019will}. To address the issues like NISQ, the classic-quantum split pattern slices the overall quantum software or application into classical modules (pre/post-processing) and quantum modules (quantum computation) that result in hybrid applications\cite{R12_valencia2022quantum}. One of the prime examples of classic-quantum split patterns is Shor’s algorithm which involves quantum computations for finding the prime factors of an integer with its application in computer security and cryptography. 
Quantum service-orientation when viewed from a utility computing perspective can minimise the quantum divide, a prevailing issue highlighted at the World Economic Forum 2023, between states/entities that own or lack QC systems, technologies, and infrastructures \cite{R24_WorldEconomicForum}. 

\subsection{Research Method for the SMS}\label{sec:method}
We now discuss the research method, driven by three phases as illustrated in Figure \ref{Researchmethod}, based on the guidelines to conduct the SMS \cite{R16_petersen2008systematic}.

\subsubsection{\textsf{Phase I} – Specifying the Research Questions}
Research questions (RQs) are fundamental to conducting the SMS and documenting the results. We outlined two RQs for this SMS.


\begin{tcolorbox} [sharp corners, boxrule=0.1mm,]
\small
\textbf{RQ-1}: What solutions are reported in the literature to support the development of quantum computing as a service?
\end{tcolorbox}

\textbf{Objective(s)} - To investigate state-of-the-art in terms of existing solutions that enable or enhance QCaaS computing. A multi-perspective analysis can reveal the functionality offered by the solutions, modeling languages and patterns to design the solutions, and programming technologies along with deployment platforms to implement and operationalise the solutions. 


\begin{tcolorbox} [sharp corners, boxrule=0.1mm,]
\small
\textbf{RQ-2}: What are the emerging trends of research on quantum computing as a service?
\end{tcolorbox}

\textbf{Objective(s)} - To identify and discuss the emerging trends that can help pinpoint the prevalent challenges and their solutions as dimensions of potentially futuristic research on QCaaS. The emerging trends can help to provide a road map for progressing research and development on QCaaS.

\vspace{0.5em}

\begin{table*}[b!] 
\scriptsize
\caption{Criteria for Screening and Qualitative Assessment of Selected Studies}
\begin{center}
{\tiny}
\begin{tabular}{|l|}
\hline
\rowcolor[HTML]{F2F2F2} 
\multicolumn{1}{|c|}{\cellcolor[HTML]{F2F2F2}\textbf{Study Selection Step I -   Screening of Identified Studies}}                                                                                \\ \hline
S1 - The study does not discuss any   solution or proposal for quantum computing as a service                                                                                                    \\ \hline
S2 - The study is not reported in  English                                                                                                                                                      \\ \hline
S3 - The study is a duplicate study. Duplicate studies are studies with overlapping contents, e.g., a conference paper extended as a journal  article.                                           \\ \hline
S4 - The study is a secondary  study/survey paper                                                                                                                                               \\ \hline
\rowcolor[HTML]{BDD6EE} 
\begin{tabular}[c]{@{}l@{}}Exclude the study if the answer to any of the criteria in Step I (S1 - S4) results in Yes, otherwise, \\ Include the study for quality assessment in Step II\end{tabular}  \\ \hline
\rowcolor[HTML]{F2F2F2} 

\multicolumn{1}{|c|}{\cellcolor[HTML]{F2F2F2}\textbf{Study Selection Step II - Quality Assessment of the Identified Studies}}                                                                    \\ \hline
Q1 - Study objectives and  Contributions are clear? {[}Yes = 1, Partially = 0.5, No = 0{]}                                                                                                      \\ \hline
Q2 - Research method to conduct  the study is reported {[}Yes = 1, Partially = 0.5, No = 0{]}                                                                                                   \\ \hline
Q3 - Design and/or implementation  details of solution are provided {[}Yes = 1, Partially = 0.5, No = 0{]}                                                                                      \\ \hline
Q4 - Details for   Experiments/Evaluation/Demonstration of Solution are provided {[}Yes = 1,  Partially = 0.5, No = 0{]}                                                                        \\ \hline
Q5 - Study limitations and needs  for future research are discussed {[}Yes = 1, Partially = 0.5, No = 0{]}                                                                                      \\ \hline
\rowcolor[HTML]{BDD6EE} 
\begin{tabular}[c]{@{}l@{}}Exclude the study that has a quality assessment score (Q1 – Q5) less than 2.0, otherwise  \\ Include the study for review and data extraction in Table 2\end{tabular} \\ \hline
\end{tabular}
\end{center}
\label{tab:qualitycriteria}
\end{table*}
\subsubsection{\textsf{Phase II} – Identifying and Selecting the Literature for SMS}

\begin{figure}[]
 \centering
 \includegraphics[scale=0.83]{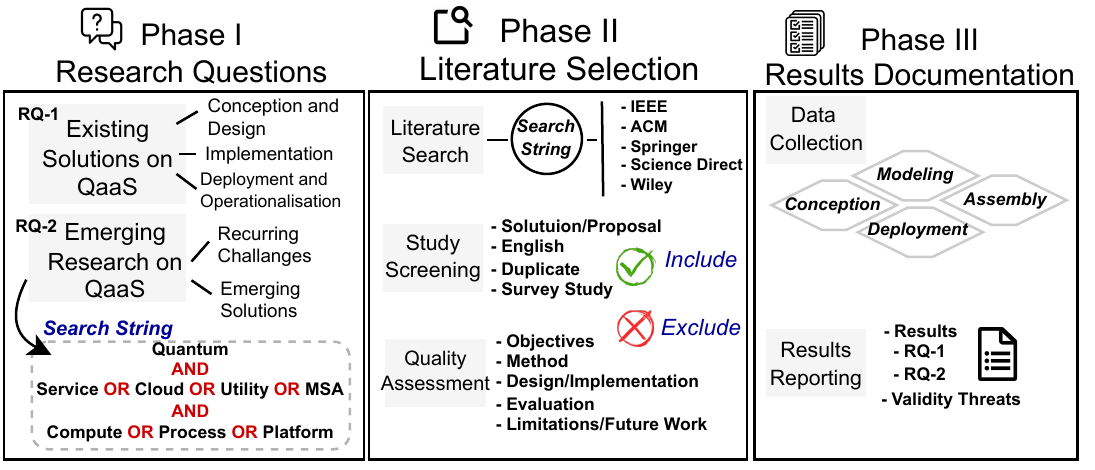} 
 	\caption{Overview of the Research Method}
	\label{Researchmethod}
\end{figure}

Based on the guidelines for literature search, we formulated a generic string to be executed on prominent Electronic Data Sources (EDS) \cite{R17_chen2010towards} including IEEE Xplore, ACM Digital Library, Springer Link, Science Direct, Springer Link, and Wiley Online Library. Google Scholar was used as a complementary EDS to ensure that we did not miss any relevant study for selection. The search string presented in Figure \ref{Researchmethod} is generic that combines logical operations \textsc{AND}, \textsc{OR}) to compose the key terms (e.g., Quantum \textsc{AND} Service \textsc{OR} Cloud), customized for each EDS individually. Customised search strings are provided as part of the SMS protocol \cite{18_SMSProtocol}. We conducted a pilot search to assess the need for any customisation to the search string(s) or any filters applied on specific EDS to avoid an exhaustive search resulting in a significant number of unrelated studies. For example, we limited our search on IEEE Xplore from `Full Text \& Metadata’ to `Document Title’ as searching for our defined key terms in full text and metadata yielded a significant amount of irrelevant studies (e.g., cloud services, quantum hardware). 

\textbf{Screening and Quality Assessment:} By executing the customised search strings on five selected EDS, the SMS process retrieved a total of 55 potentially relevant studies. To complement the automated search on EDS, we applied the forward snowballing process [17], as a manual effort. The forward snowballing approach involves looking up the references or bibliography sections of 55 studies, referred to as the seed set in snowballing, to see if any relevant cited literature can be found.  The forward snowballing helped us to identify a total of 13 studies resulting in a total of 68 studies (55: EDS and 13: snow-balling). To assess and select the studies for review, we performed study screening based on criteria (Step 1: S1 – S4) in Table \ref{tab:qualitycriteria}. Most of the studies identified during the snowballing failed the screening criteria S3 and S4 which means either the studies were duplicate studies or secondary/survey studies that cannot be included in the review. Based on the screening of identified studies, more specifically reading through the titles, abstracts, and conclusions we shortlisted a total of 11 potentially relevant studies to be qualitatively assessed (Step II: Q1 – Q5) for their inclusion in the review for SMS. Based on the quality assessment, we excluded 2 studies to finally select a total of 9 studies to be included in the review. The list of selected studies for the SMS is provided in \textbf{Appendix A}. 

\vspace{0.5em}
\subsubsection{\textsf{Phase III} – Documenting the Results}
To document the results, i.e., answering RQs objectively, we extracted the data from the selected studies in Appendix A and documented it using a structured format, having seven criteria, in Table \ref{tab:criteria}. The criteria focus on conceptualising, designing, developing, and deploying quantum software services by following the IBM SOA foundation life cycle (SOA life cycle for short) \cite{R20_keen2006patterns}. To contextualise QCaaS from Figure \ref{Fig-1:Context} and to ensure fine-grained analysis of SMS data, we have divided the `Model' activity from SOA life cycle into two activities namely Conception and Model to distinguish between functional needs (conception) and representation (modeling) of quantum service design. Model represents the conception as the design specification of functional needs for quantum services. We do not have the `Manage' phase from SOA life cycle as we could not find any evidence in the literature that supports identity, compliance, and business metrics management of quantum services.

\begin{itemize}
\item \textbf{Conception} as the initial activity in the service life cycle aims to conceptualise the functional aspects of quantum services by capturing the details of required functionality, i.e., functional requirements. Conception aims to identify: \textit{what are the business needs of a quantum service?}

\item \textbf{Model} activity aims to translate the conception into a design that acts as a blueprint for implementing quantum services. Model focuses on: \textit{how to represent the conception as the design of the solution?}

\item \textbf{Assemble} focuses on implementing the design to produce concrete, i.e., executable specification of quantum services. Assemble aims to address: \textit{how to implement the design as executable services?}

\item \textbf{Deploy} as the last activity aims to deploy the assembled solution for operationalisation and usage of quantum services. Deploy focuses on: \textit{what platforms can be used to deploy the assembled (implemented) solution?}
\end{itemize}

\vspace{0.5em}

\subsubsection{Threats to the Validity of SMS}
Systematic literature reviews and mapping studies are prone to a number of validity threats that refer to deviation, limitations, or invalidation of study results when applied to a theoretical or practical context. 
\textbf{Construct validity} of the SMS corresponds to the rigor of study protocol and methodological details to extract, analyze, and synthesise the data to objectively answer the RQs and present the data systematically. To avoid this threat, i.e., avoiding the bias in data extraction and documentation, we applied well-practiced guidelines \cite{R16_petersen2008systematic, R17_chen2010towards}, derived the search strings (Figure \ref{Researchmethod}), and devised a structured format (Table \ref{tab:criteria}) to collect and present the data consistently. \textbf{Internal validity} examines SMS design, conduct, and analysis to answer the RQs without bias. To minimize this threat, we synthesized the data based on the well-known IBM SOA life cycle \cite{R20_keen2006patterns} that structures the results into fine-grained life cycle activities. We documented the results while performing a quality assessment (Table \ref{tab:qualitycriteria}) and a well-defined service life cycle template. \textbf{External validity} of the SMS refers to the extent to which the findings of study can be generalised/externalised to research and development projects. It is challenging to foresee and outline the predictive implications of the study results. We have outlined the implications and generalization of study findings (Table \ref{tab:criteria}, Figure \ref{Results}, Figure \ref{FutureResearch}) can provide the basis for creating a reference architecture for QCaaS as future work. The documented results are discussed in Section \ref{sec:RQ1} (RQ-1) and Section \ref{sec:RQ2} (RQ-2).

\section{Engineering Software for QCaaS (RQ-1)}
\label{sec:RQ1}
We now discuss the existing solutions, reported in the literature, that support the development of quantum services to operationalise QCaaS solutions. The data extracted from the selected studies is presented in Table \ref{tab:criteria} and visualised in Figure \ref{Results}. Table \ref{tab:criteria} can be viewed as a catalogue that organises a summary of the core findings to answer RQ-1 based on the four activities of the SOA life cycle (Phase III, Figure \ref{Researchmethod}). 

\vspace{0.5em}

\textbf{Illustrative Example}: Figure \ref{Results}a) exemplifies the service life cycle with \textit{functional aspects} that requires a quantum service to compute the prime factors of an integer. Functional aspects need \textit{modeling} and that uses Unified Modeling Language (UML) component diagram \cite{R21_perez2020towards} as the modeling notation to specify computational elements (components) and their interconnections (connectors) in a service. The Classic-Quantum split pattern \cite{R12_valencia2022quantum} is applied to slice the functionality between a classical computer (pre-/post-processing, e.g., \textsf{Gen\_Num} component) and a quantum computer (prime factorization, \textsf{Factorize} component). The model acts as a blue-print to support \textit{assembly} of a services using a programming language (Qiskit code snippet) that converts the design specifications of a service to its executable specifications. Finally, \textit{deployment} activity is shown as UML deployment diagram to configure the assembled service on a quantum computer provided by the quantum vendor (Amazon Braket). 

\subsection{\textbf{Conception}: \textsf{Functional Aspects}} 
\begin{figure*}[ht]
 \centering
 \includegraphics[scale =0.75]{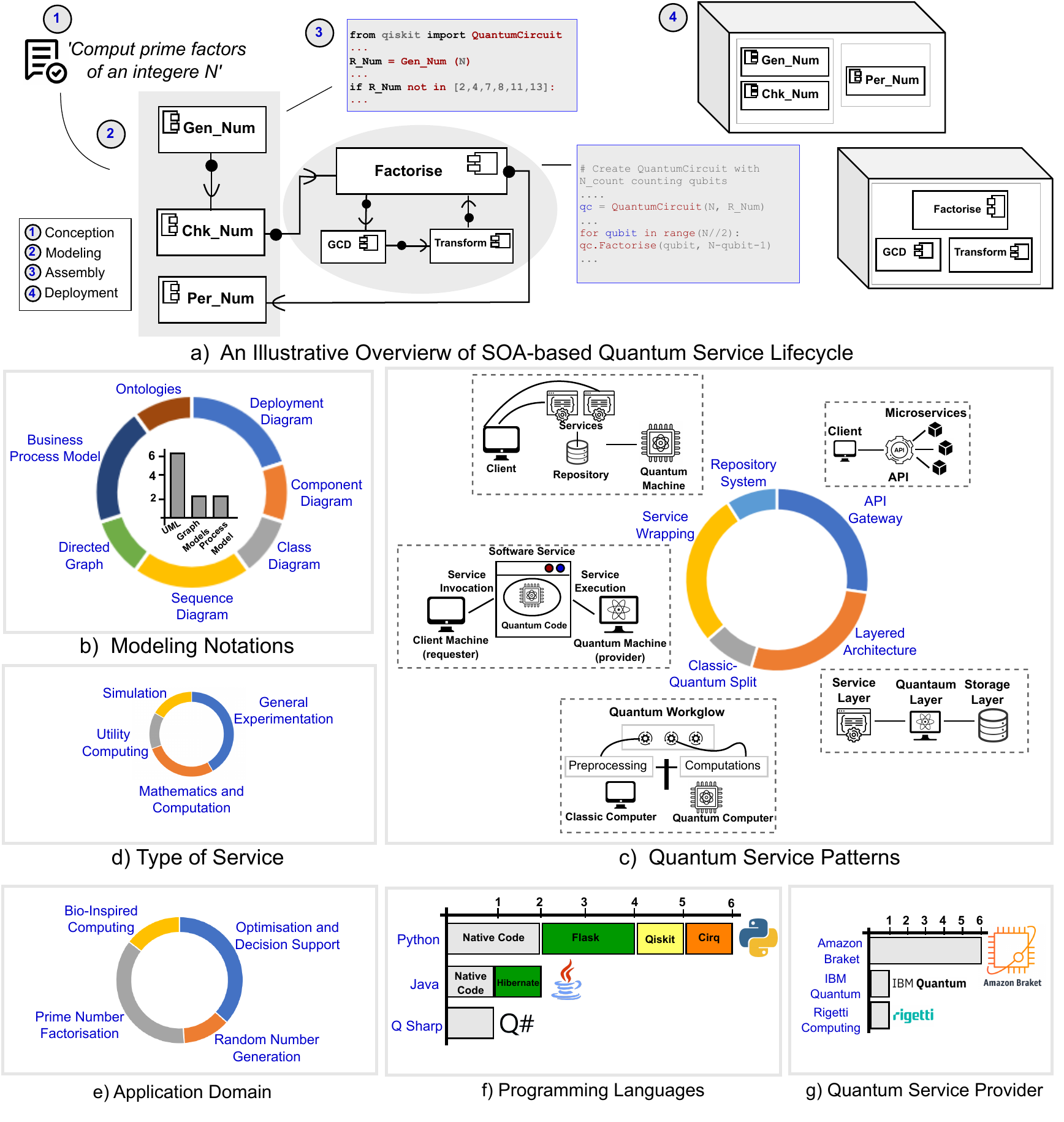} 
 	\caption{An Overview of the Mapping Study Results}
	\label{Results}
\end{figure*}

During the conception activity, functional aspects of a service relate to identifying and outlining the functionality to be offered by a quantum service. The functionality can be achieved via service execution on a quantum computer (provider), whereas the service is developed or invoked by the user (requester). Figure \ref{Fig-1:Context} shows that, in order to exploit QC platforms for quantum functionality, service requesters can develop/discover new and/or available services, such as quantum simulation or quantum cryptography using the SOA patterns \cite{R7_bouguettaya2017service}. In the SMS, we identified a multitude of functional aspects for quantum services and organized them into five categories namely \textit{experimental}, \textit{service delivery}, \textit{number crunching}, \textit{data searching}, and \textit{natural computing} as shown in Table \ref{tab:criteria}. For example, Figure \ref{Results}a) highlights the generic functional aspect of number crunching that involves prime factorisation of integers for Shor's algorithm \cite{R1_ali2022software}\cite{R2_harrow2017quantum}.  

The diversity of existing functional aspects is proportional to the capability of the current era of quantum computers that is considred as limited due to a number of factors such as simplistic quantum circuitry (e.g., less QuBits/QuGates) and quantum errors (e.g., NISQ) \cite{R4_dyakonov2019will}. Quantum software services that support functional aspects of QC are merely capable of checking the status of quantum circuits (experimental) or generation of random numbers using quantum hardware (number crunching). Functional aspects reflect only a partial view of system design in terms of offered functionality that should not overlook the non-functional or quality aspects of the QCaaS solutions. For example, resource efficiency in terms of utilizing minimal available QuBits to generate prime factors can ensure the required functionality and desired quality (e.g., service efficiency, execution performance) of QCaaS.  

\begin{tcolorbox} [sharp corners, boxrule=0.1mm,]
\faEdit \scriptsize{~\textsf{\textbf{Functional aspects} of quantum services, in general, are rather limited to basic quantum experimentation and numerical processing. The limitation reflects the existing capabilities of QCs and consequently the offered services by quantum vendors. Investigating the non-functional aspects can help outline the quantum significant requirements (QSRs) in terms of required functionality and desired quality of the service that currently lacks in the existing literature.}}
\end{tcolorbox}

\subsection{\textbf{Modeling}: \textsf{Notations and Patterns}} 

During service modeling, modeling notations, such as Q-UML or ontological models can help create a blueprint for the implementation of functional aspects \cite{R21_perez2020towards}. Patterns can complement the notations by providing reusable knowledge and design rationale to architect the quantum services \cite{R10_garcia2021quantum} \cite{R12_valencia2022quantum}.

\begin{table*}[]
\caption{Data Extracted for SMS from Reviewed Studies (SOA lifecycle activities [20])}
\begin{centering}
{\tiny
\begin{tabular}{|c|c|cc|cc|c|}
\hline
\multicolumn{7}{|c|}{\includegraphics[scale=0.6]{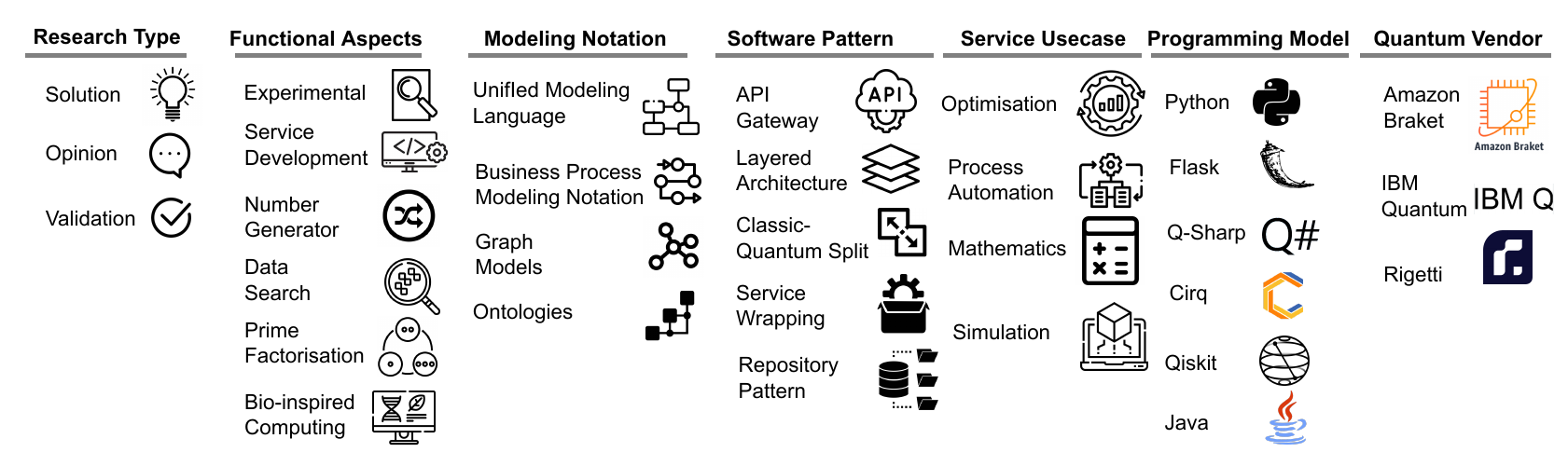}} \\ \hline
\multicolumn{1}{|l|}{} & \multicolumn{1}{c|}{\cellcolor[HTML]{BDD6EE}\textbf{Conception}} & \multicolumn{2}{c|}{\cellcolor[HTML]{BDD6EE}\textbf{Modeling}} & \multicolumn{2}{c|}{\cellcolor[HTML]{BDD6EE}\textbf{Assembly}} & \multicolumn{1}{c|}{\cellcolor[HTML]{BDD6EE}\textbf{Deployment}} \\ \cline{2-7}
\multirow{-2}{*}{\textbf{\begin{tabular}[c]{@{}c@{}}Study ID \\ \& \\ Research Type\end{tabular}}} &
  \cellcolor[HTML]{F2F2F2}\begin{tabular}[c]{@{}c@{}}Functional \\ Aspects\end{tabular} &
  \multicolumn{1}{c|}{\cellcolor[HTML]{F2F2F2}\begin{tabular}[c]{@{}c@{}}Modelling \\ Notation\end{tabular}} &
  \cellcolor[HTML]{F2F2F2}\begin{tabular}[c]{@{}c@{}}Software\\ Pattern\end{tabular} &
  \multicolumn{1}{c|}{\cellcolor[HTML]{F2F2F2}\begin{tabular}[c]{@{}c@{}}Service \\ Use case\end{tabular}} &
  \cellcolor[HTML]{F2F2F2}\begin{tabular}[c]{@{}c@{}}Service\\ Programming\end{tabular} &
  \cellcolor[HTML]{F2F2F2}\begin{tabular}[c]{@{}c@{}}Quantum\\ Platform/\\ Vendor\end{tabular} \\ \hline
\rowcolor[HTML]{FFFFFF}

\begin{tabular}[c]{@{}c@{}}{\includegraphics[scale=0.11]{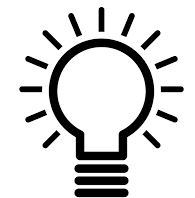}} \\ {[}S1{]}\end{tabular} &
  \begin{tabular}[c]{@{}c@{}}{\includegraphics[scale=0.04]{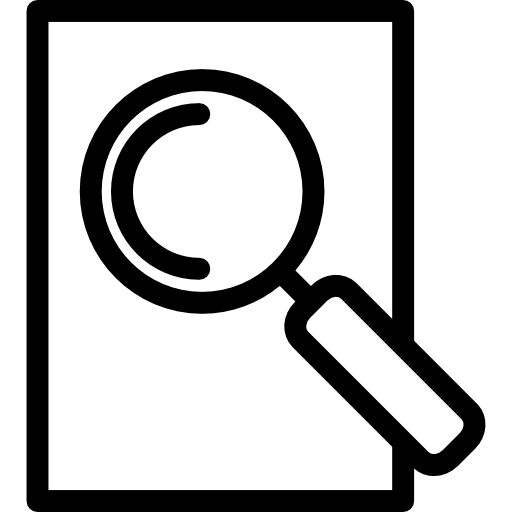}}\\    Quantum   Service \\ Delivery\end{tabular} &
  \multicolumn{1}{c|}{\cellcolor[HTML]{FFFFFF}\begin{tabular}[c]{@{}c@{}}{\includegraphics[scale=0.04]{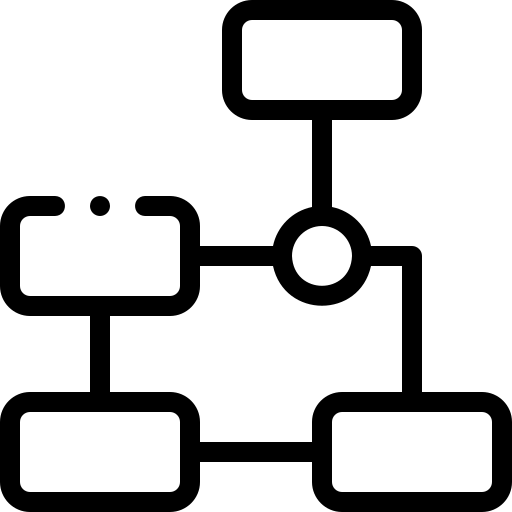}}   \\ Deployment\\ Diagram\end{tabular}} &
  \begin{tabular}[c]{@{}c@{}}{\includegraphics[scale=0.04]{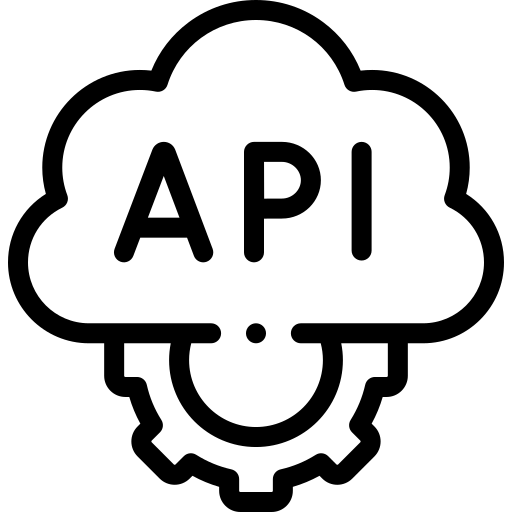}}\\ API Gateway\end{tabular} &
  \multicolumn{1}{c|}{\cellcolor[HTML]{FFFFFF}\begin{tabular}[c]{@{}c@{}}{\includegraphics[scale=0.04]{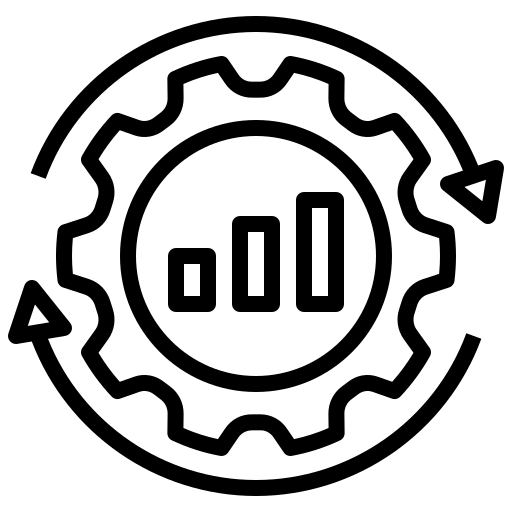}} \\ Optimal Service \\ Provider\end{tabular}} &

  \begin{tabular}[c]{@{}c@{}}{\includegraphics[scale=0.15]{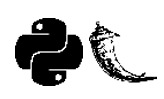}} \\ Python, Flask \end{tabular} &
  
  \begin{tabular}[c]{@{}c@{}}{\includegraphics[scale=0.20]{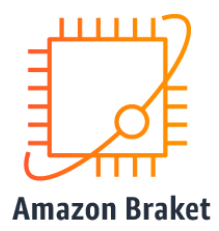}}\end{tabular} \\ \hline
\rowcolor[HTML]{FFFFFF} 
\begin{tabular}[c]{@{}c@{}}{\includegraphics[scale=0.04]{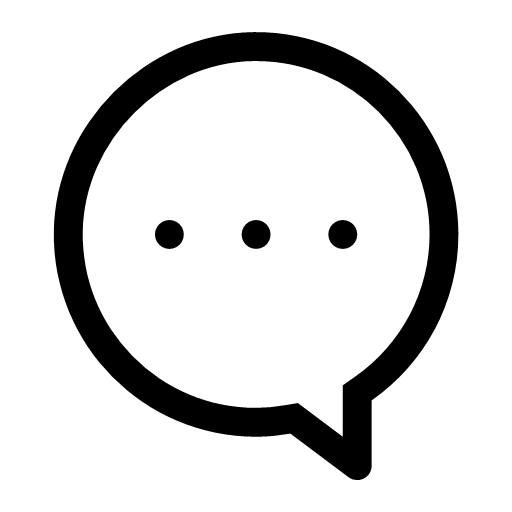}}\\   {[}S2{]}\end{tabular} &
  \begin{tabular}[c]{@{}c@{}}{\includegraphics[scale=0.07]{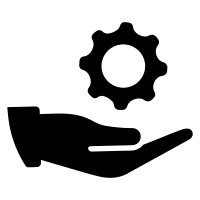}}   \\ Enterprise  Services \\ Development\end{tabular} &
  \multicolumn{1}{c|}{\cellcolor[HTML]{FFFFFF}\begin{tabular}[c]{@{}c@{}}{\includegraphics[scale=0.04]{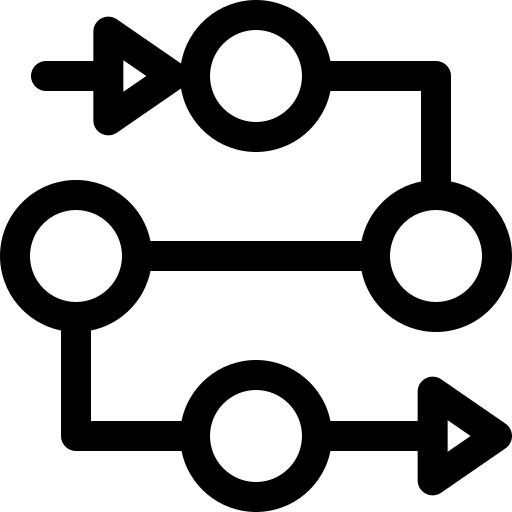}}\\   Business Process\end{tabular}} &

  \begin{tabular}[c]{@{}c@{}}{\includegraphics[scale=0.07]{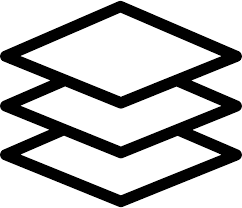}}  \\ Layered\\ Architecture\end{tabular} &
  
  \multicolumn{1}{c|}{\cellcolor[HTML]{FFFFFF}\begin{tabular}[c]{@{}c@{}}{\includegraphics[scale=0.04]{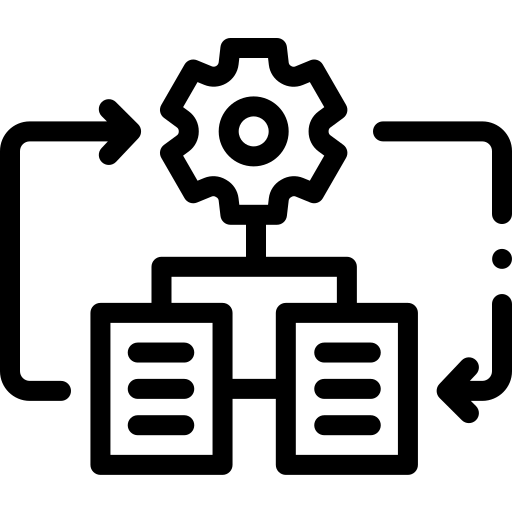}}   \\ Process Automation\end{tabular}} &
  
  \begin{tabular}[c]{@{}c@{}}{\includegraphics[scale=0.07]{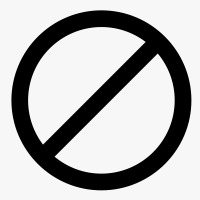}}\\No Evidence \end{tabular} &
  \begin{tabular}[c]{@{}c@{}}{\includegraphics[scale=0.07]{Images/icons/Noevidence.png}}\\No Evidence \end{tabular} \\ \hline
\rowcolor[HTML]{FFFFFF} 
\begin{tabular}[c]{@{}c@{}}{\includegraphics[scale=0.11]{Images/icons/1.jpg}}\\   {[}S3{]}\end{tabular} &
  \begin{tabular}[c]{@{}c@{}}{\includegraphics[scale=0.11]{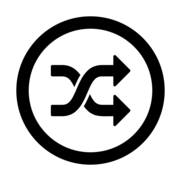}}   \\ Quantum Random \\ Number Generation\\ Quantum Search Algo\end{tabular} &
  \multicolumn{1}{c|}{\cellcolor[HTML]{FFFFFF}\begin{tabular}[c]{@{}c@{}}{\includegraphics[scale=0.04]{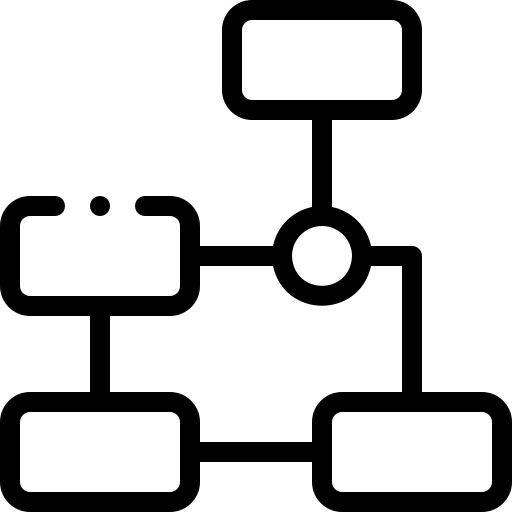}}   \\ Class, Sequence \\ Diagram\end{tabular}} &

  \begin{tabular}[c]{@{}c@{}}{\includegraphics[scale=0.07]{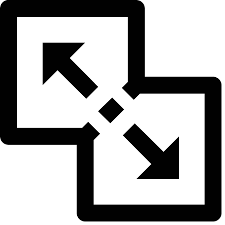}}\\Classic-
Quantum Split\end{tabular} &
  
  \multicolumn{1}{c|}{\cellcolor[HTML]{FFFFFF}\begin{tabular}[c]{@{}c@{}}{\includegraphics[scale=0.07]{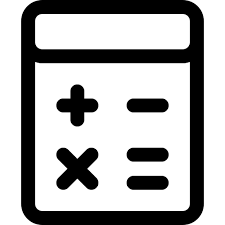}} \\ Mathematics\end{tabular}} &
  
  \begin{tabular}[c]{@{}c@{}}{\includegraphics[scale=0.11]{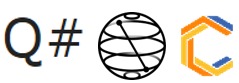}} \\Q sharp\end{tabular} &

  \begin{tabular}[c]{@{}c@{}}{\includegraphics[scale=0.11]{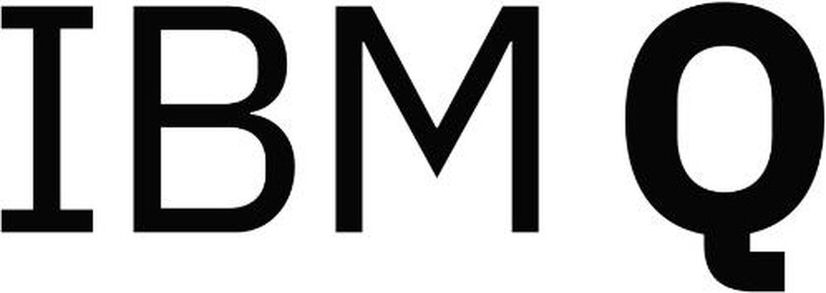}} \\IBM Quantum\end{tabular} \\ \hline
\rowcolor[HTML]{FFFFFF} 
\begin{tabular}[c]{@{}c@{}}{\includegraphics[scale=0.04]{Images/icons/opinion.png}} \\ {[}S4{]}\end{tabular} &
  
  \begin{tabular}[c]{@{}c@{}}{\includegraphics[scale=0.11]{Images/icons/numb.png}}\\ Integer Factorisation\end{tabular} &
  
  \multicolumn{1}{c|}{\cellcolor[HTML]{FFFFFF}\begin{tabular}[c]{@{}c@{}}{\includegraphics[scale=0.05]{Images/icons/UMLS.png}}\\ Deployment Diagram\end{tabular}} &
  \begin{tabular}[c]{@{}c@{}}{\includegraphics[scale=0.11]{Images/icons/1.jpg}} \\ Solution\end{tabular} &
  
  \multicolumn{1}{c|}{\cellcolor[HTML]{FFFFFF}\begin{tabular}[c]{@{}c@{}}{\includegraphics[scale=0.07]{Images/icons/maths.png}} \\ Mathematics\end{tabular}} &
  
  \begin{tabular}[c]{@{}c@{}}{\includegraphics[scale=0.005]{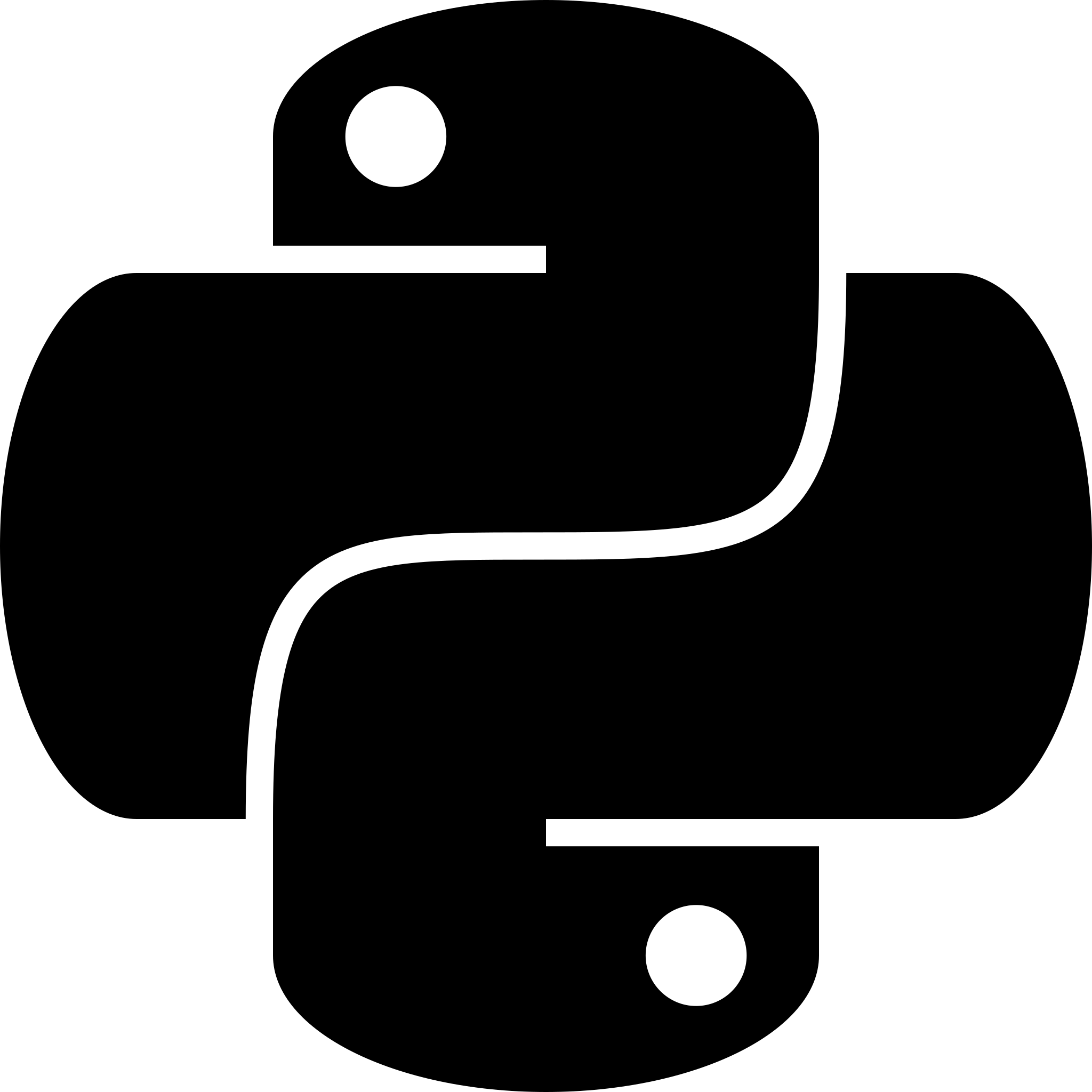}}\\Python\end{tabular} &
  
  \begin{tabular}[c]{@{}c@{}}{\includegraphics[scale=0.20]{Images/icons/amazon.png}}\end{tabular} \\ \hline
  
  \rowcolor[HTML]{FFFFFF} 
\begin{tabular}[c]{@{}c@{}}{\includegraphics[scale=0.01]{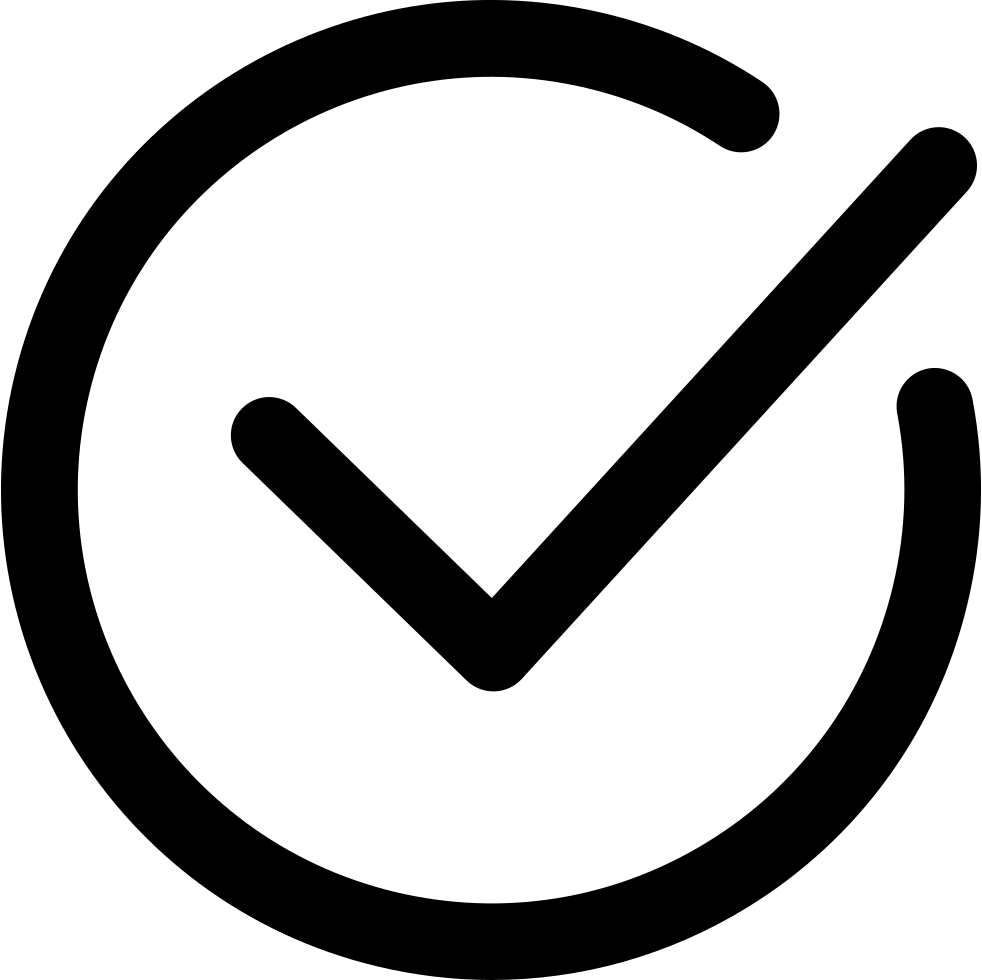}}  \\ {[}S5{]}\end{tabular} &
  \begin{tabular}[c]{@{}c@{}}{\includegraphics[scale=0.07]{Images/icons/ServiceDelivery.png}}\\ Experimental Quantum\\   Service Computing)\end{tabular} &
  
  \multicolumn{1}{c|}{\cellcolor[HTML]{FFFFFF}\begin{tabular}[c]{@{}c@{}}{\includegraphics[scale=0.07]{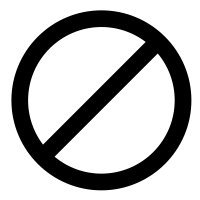}}\\No Evidence\end{tabular}} &
  
  \begin{tabular}[c]{@{}c@{}}{\includegraphics[scale=0.09]{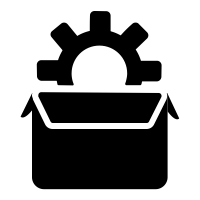}}\\Service  
Wrapping \end{tabular} &
  
  \multicolumn{1}{c|}{\cellcolor[HTML]{FFFFFF}\begin{tabular}[c]{@{}c@{}}{\includegraphics[scale=0.04]{Images/icons/oPTIMISATIOPN.png}}\\Optimisation\end{tabular}} &

  \begin{tabular}[c]{@{}c@{}}{\includegraphics[scale=0.15]{Images/icons/flask1.png}}\\Python, Flask\end{tabular} &
  
  \begin{tabular}[c]{@{}c@{}}{\includegraphics[scale=0.20]{Images/icons/amazon.png}}\end{tabular} \\ \hline

\rowcolor[HTML]{FFFFFF} 
\begin{tabular}[c]{@{}c@{}}{\includegraphics[scale=0.11]{Images/icons/1.jpg}}\\   {[}S6{]}\end{tabular} &
  \begin{tabular}[c]{@{}c@{}}{\includegraphics[scale=0.07]{Images/icons/ServiceDelivery.png}}\\ Experimental Services\\ Algorithm\end{tabular} &
  \multicolumn{1}{c|}{\cellcolor[HTML]{FFFFFF}\begin{tabular}[c]{@{}c@{}}{\includegraphics[scale=0.04]{Images/icons/UML.png}}\\ Sequence Diagrams\end{tabular}} &
  
  \begin{tabular}[c]{@{}c@{}}{\includegraphics[scale=0.04]{Images/icons/API.png}}\\ API Gateway\end{tabular} &

  \multicolumn{1}{c|}{\cellcolor[HTML]{FFFFFF}\begin{tabular}[c]{@{}c@{}}{\includegraphics[scale=0.07]{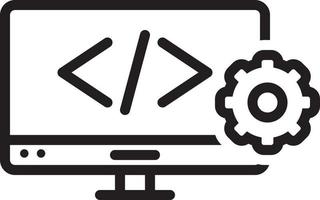}}\\ Algorithm as \\ a Service\end{tabular}} &
 
  \begin{tabular}[c]{@{}c@{}}{\includegraphics[scale=0.005]{Images/icons/Python_icon_black_and_white.png}}\\Python \end{tabular} &
  
  \begin{tabular}[c]{@{}c@{}}{\includegraphics[scale=0.07]{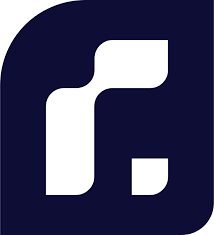}} \\ Rigetti\end{tabular} \\ \hline

\rowcolor[HTML]{FFFFFF} 
\begin{tabular}[c]{@{}c@{}}{\includegraphics[scale=0.11]{Images/icons/1.jpg}}\\   {[}S7{]}\end{tabular} &
  \begin{tabular}[c]{@{}c@{}}{\includegraphics[scale=0.07]{Images/icons/numb.png}}   \\ Integer Factorisation\end{tabular} &
  
  \multicolumn{1}{c|}{\cellcolor[HTML]{FFFFFF}\begin{tabular}[c]{@{}c@{}}{\includegraphics[scale=0.15]{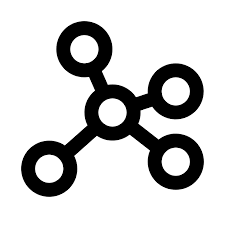}}\\Directed Graph\end{tabular}} &
  
  \begin{tabular}[c]{@{}c@{}}{\includegraphics[scale=0.04]{Images/icons/API.png}}\\ API Gateway\end{tabular} &
  
  \multicolumn{1}{c|}{\cellcolor[HTML]{FFFFFF}\begin{tabular}[c]{@{}c@{}}{\includegraphics[scale=0.04]{Images/icons/oPTIMISATIOPN.png}} \\Optimisation \end{tabular}} &
  
  \begin{tabular}[c]{@{}c@{}}{\includegraphics[scale=0.07]{Images/icons/Noevidence.png}}\\No Evidence\end{tabular} &
  
  \begin{tabular}[c]{@{}c@{}}{\includegraphics[scale=0.20]{Images/icons/amazon.png}}\end{tabular} \\ \hline

\rowcolor[HTML]{FFFFFF} 
\begin{tabular}[c]{@{}c@{}}{\includegraphics[scale=0.04]{Images/icons/opinion.png}}   \\ {[}S8{]}\end{tabular} &

  \begin{tabular}[c]{@{}c@{}}{\includegraphics[scale=0.03]{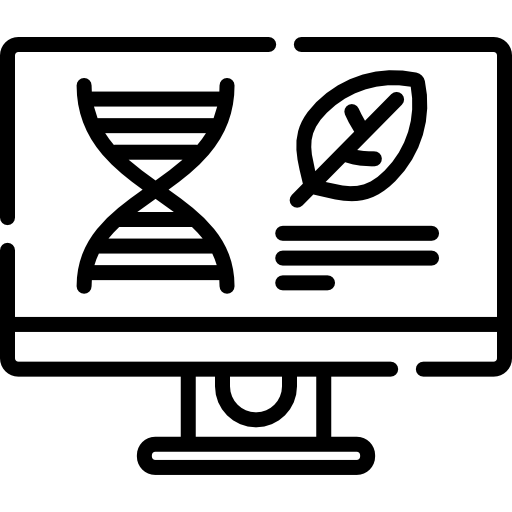}}\\Bio-inspired\\
Computing\end{tabular} &
  
  \multicolumn{1}{c|}{\cellcolor[HTML]{FFFFFF}\begin{tabular}[c]{@{}c@{}}{\includegraphics[scale=0.07]{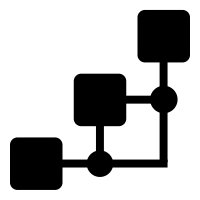}}\\Ontologies\end{tabular}} &
  
  \begin{tabular}[c]{@{}c@{}}{\includegraphics[scale=0.09]{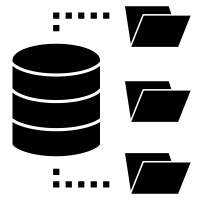}}\\Repository \\
Pattern\end{tabular} &
  
  \multicolumn{1}{c|}{\cellcolor[HTML]{FFFFFF}\begin{tabular}[c]{@{}c@{}}{\includegraphics[scale=0.04]{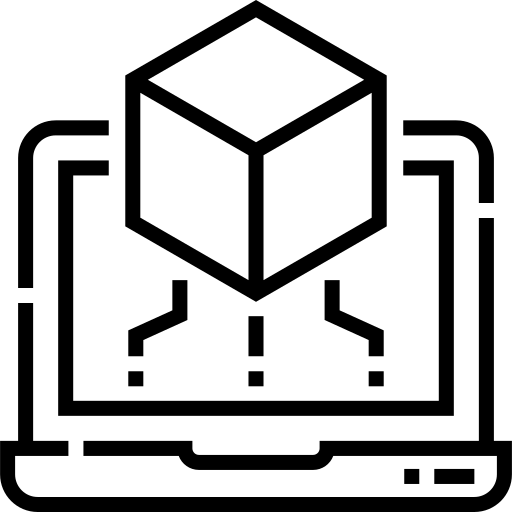}} \\Simulation\end{tabular}} &
  
  \begin{tabular}[c]{@{}c@{}}{\includegraphics[scale=0.07]{Images/icons/Noevidence.png}}\\No Evidence\end{tabular} &
  
  \begin{tabular}[c]{@{}c@{}}{\includegraphics[scale=0.07]{Images/icons/Noevidence.png}}\\No Evidence\end{tabular} \\ \hline

\rowcolor[HTML]{FFFFFF} 
\begin{tabular}[c]{@{}c@{}}{\includegraphics[scale=0.04]{Images/icons/opinion.png}}   \\ {[}S9{]}\end{tabular} &
 
  \begin{tabular}[c]{@{}c@{}}{\includegraphics[scale=0.11]{Images/icons/numb.png}} \\Integer \\Factorisation \end{tabular} &
  
  \multicolumn{1}{c|}{\cellcolor[HTML]{FFFFFF}\begin{tabular}[c]{@{}c@{}}{\includegraphics[scale=0.07]{Images/icons/Noevidence.png}}\\No Evidence\end{tabular}} &
  
  \begin{tabular}[c]{@{}c@{}}{\includegraphics[scale=0.09]{Images/icons/servicewrapper.png}} \\Service  
Wrapping\end{tabular} &
  
  \multicolumn{1}{c|}{\cellcolor[HTML]{FFFFFF}\begin{tabular}[c]{@{}c@{}}{\includegraphics[scale=0.07]{Images/icons/Noevidence.png}}\\No Evidence\end{tabular}} &
  
  \begin{tabular}[c]{@{}c@{}}{\includegraphics[scale=0.005]{Images/icons/Python_icon_black_and_white.png}}\\Python\end{tabular} &
  \begin{tabular}[c]{@{}c@{}}{\includegraphics[scale=0.20]{Images/icons/amazon.png}}\end{tabular} \\ \hline
\end{tabular}}
\par\end{centering}
\label{tab:criteria}
\end{table*}

\textit{Modeling notations} are fundamental to the creation, maintenance, and evolution of models such as ontological structures and graph-based diagrams that provide a visual representation, whereas architectural description languages support a textual specification for software-intensive systems \cite{R21_perez2020towards}. Recent trends in software engineering that promote model-driven and low-code application development have resulted in transitioning developers’ focus from coding to software modeling for implementation \cite{R22_raymer2019us}. Low code application development process leverages the principle and practices of model-driven engineering to utilise model(s) as first-class entities in software development \cite{R15_de2022software}\cite{R19_ahmadtowards}. Investigating software models and modeling notations that help create service models is essential to support model-driven perspective to QSE, consequently facilitating quantum code developers to abstract implementation-specific complexities, via model-driven QSE, while developing quantum services \cite{R19_ahmadtowards}. This SMS indicates three main types of notations to model services in QCaaS that include the Unified Modeling Language (UML), graph-based models, and process models highlighted in Figure \ref{Results}b). UML-based models are represented via a multitude of notations, such as class and component diagrams that represent the structure, while sequence and deployment diagrams represent runtime or behavioural view of QCaaS. Graph-based models contain directed graphs and ontologies, whereas process models rely on automating the business processes of an enterprise as quantum services. For example, the study [S3] reports a class diagram as a structural view of the system to represent the attributes and methods of entities (user, service provider, authentication etc.) of quantum computing as a service. 
UML diagrams and profiles represent the status-quo in software modeling and are seen as the de-facto notation in the software and service ' community to model classical software with growing adoption in QSE \cite{R21_perez2020towards}.

\textit{Design patterns} represent a concentrated wisdom of software designers that can be leveraged to address design and implementation issues, addressing functionality and quality,   effectively and efficiently. Considering a lack of professional expertise in QSE (e.g., quantum domain engineers, quantum algorithm designers, quantum software architects etc.) patterns as artifacts of reuse can help novice developers during quantum software development to rely on existing best practices \cite{R1_ali2022software}\cite{R10_garcia2021quantum}. This SMS highlights that the literature on QCaaS  reports five patterns, namely the \textit{API Gateway}, \textit{Layered Architecture}, \textit{Classic-Quantum Split}, \textit{Service Wrapping}, and \textit{Repository Pattern}. Figure \ref{Results}c) depicts pattern thumbnails as an abstract view of the identified patterns. Patterns are generally documented as templates or pattern languages, here we only focus on overviewing the reported patterns for QCaaS, while details for pattern representation and documentation can be found in \cite{R19_ahmadtowards}. The Classic-Quantum Split pattern \cite{R12_valencia2022quantum} is a quantum version of the Splitter pattern, driven by quantum workflow, that splits computation tasks into tasks that can be generated and executed on classical machines (e.g., random number generation) and tasks that can be executed on quantum machines (e.g., prime factorisation). The pattern aims to address issues like NISQ by splitting quantum software into classical and quantum parts as a hybrid application \cite{R2_harrow2017quantum}\cite{R4_dyakonov2019will}. 

\begin{tcolorbox} [sharp corners, boxrule=0.1mm,]
\faEdit \scriptsize{~\textsf{\textbf{Modeling notations} can assist software engineers to transit their focus from implementation towards design perspective. Modeling can incrementally transform functional aspects to service models leading to service implementation via model-driven engineering or low-code development. \textbf{Patterns} (classical or quantum-specific) can facilitate developers to architect and implement quantum-age software services by relying on reusable knowledge and best practices of service-orientation.}}
\end{tcolorbox}

Recently, a number of studies have focused on organising quantum software patterns as a body of knowledge in QSE, however, there is no evidence of empirically-derived methods to discover and document patterns for quantum services computing. SOA-specific patterns like API Gateway and Service Wrapping patterns can be tailored to address QCaaS solutions. There is a need for mining repositories and knowledge resources to discover reusable knowledge and best practices from quantum software development projects that can to be documented as tactics and patterns for QCaaS solutions.

\subsection{\textbf{Assembly:} \textsf{Application Domain and Programming}} 
Assembling the quantum services involves identifying the application domains and exploiting the programming languages as implementation technologies to develop executable specifications from the service model \cite{R7_bouguettaya2017service} \cite{R9_moguel2022quantum}. 

\textit{Application Domain} is also referred to as the implementation use cases or practical context to which the QCaaS solutions can be applied. For example, quantum security represents an application domain for quantum software servicing where a service can be invoked to implement cryptography protocols to generate and manage a secure quantum key \cite{R13_monroe2018quantum}. The results of this SMS indicate four application domains, namely \textit{OptimiSation}, \textit{Process Automation}, \textit{Mathematics}, and \textit{Quantum Simulation}. The application domains may impact the selection of programming languages and tools for service implementation. For example, the study [S3] uses Q\# as the programming language that can be developed and compiled in Microsoft .Net framework for executing quantum algorithms.

\textit{Service Implementation} involves programming languages that represent a system of notation or source coding scripts for implementing quantum  services to manage and operationalise QC resources \cite{R15_de2022software} \cite{R21_perez2020towards}. In recent years, a number of Quantum Programming Languages (QPLs) including but not limited to Q\# by Microsoft or Cirq by Google have emerged to provide specialized programming syntax, framework, and environments to develop, execute, and deploy quantum source code. Insights into programming languages can reveal if classical programming languages (e.g., C, Java, Python etc.) suffice for QCaaS implementation or if there is need for more specialised QPLs (Q\#, Cirq etc.). The SMS results indicate three programming languages, namely Python, Java, and Q\#, as the preferred languages to implement quantum services. Based on the details of source coding, Figure \ref{Results}f) distinguishes between native code of a language and specialised libraries/application programming interfaces (APIs) being developed using a specific language. For example, the studies [S4, S5] used native Python code to implement quantum micro-servicing for experimentation. In comparison, the study [S1] used Flask as a Web framework written in Python to develop a solution for optimal delivery of quantum services on Amazon Braket. Python is the most preferred programming language both in terms of native code as well as specialised libraries of Python that include Flask and Qiskit as open-source language frameworks and Cirq which is adopted by Google. 

\begin{tcolorbox} [sharp corners, boxrule=0.1mm,]
\faEdit \scriptsize{~\textsf{\textbf{Application domains} represent the practical context/use cases of quantum services and may impact the selection of programming languages and tools for implementation. \textbf{Programming languages} provide a system of notation for source-coding of quantum services. Classical programming languages, such as Python represent a predominant choice over QPLs to implement QCaaS due to more comprehensive documentation and familiarity of Python in service developers' community.}}
\end{tcolorbox}
    
\subsection{\textbf{Deployment:} \textsf{Quantum Platform}} 
The deployment activity supports the selection of QC platforms on which services can be deployed for their operationalisation and execution. Platform providers also referred to as quantum vendors offer computing infrastructure in terms of hardware as well as software that allows service developers to develop and/or utilise the quantum services. Deployment represents the last activity in the SOA life cycle that is represented as a UML deployment diagram in Figure \ref{Results}a). This SMS identified a total of three quantum vendors for the deployment of quantum services, namely \textit{Amazon Bracket}, \textit{IBM Quantum}, and \textit{Rigetti}. Amazon Braket (a managed Amazon Web Services (AWS)) is the most preferred platform to design, test, and run quantum algorithms. One of the reasons for selecting Amazon Braket for service deployment is that it can allow service users/developers to design their own quantum algorithms. This can be particularly handy for novice developers unfamiliar with the technicalities of quantum systems to utilise a set of pre-built algorithms, tools, and documents to develop and manage quantum services on Amazon platform. 

\begin{tcolorbox} [sharp corners, boxrule=0.1mm,]
\faEdit \scriptsize{~\textsf{\textbf{Quantum platforms} leverage cloud computing infrastructures to support quantum service-orientation. Amazon Braket is the predominant quantum vendor that can enable novice developers to exploit some pre-built algorithms, programming tools, and service documentation that lack on other quantum platforms.}}
\end{tcolorbox}

    
    

\section{Emerging Trends of Research on QaaS (RQ-2)}
\label{sec:RQ2}
\begin{figure}[]
 \centering
 \includegraphics[scale=0.7]{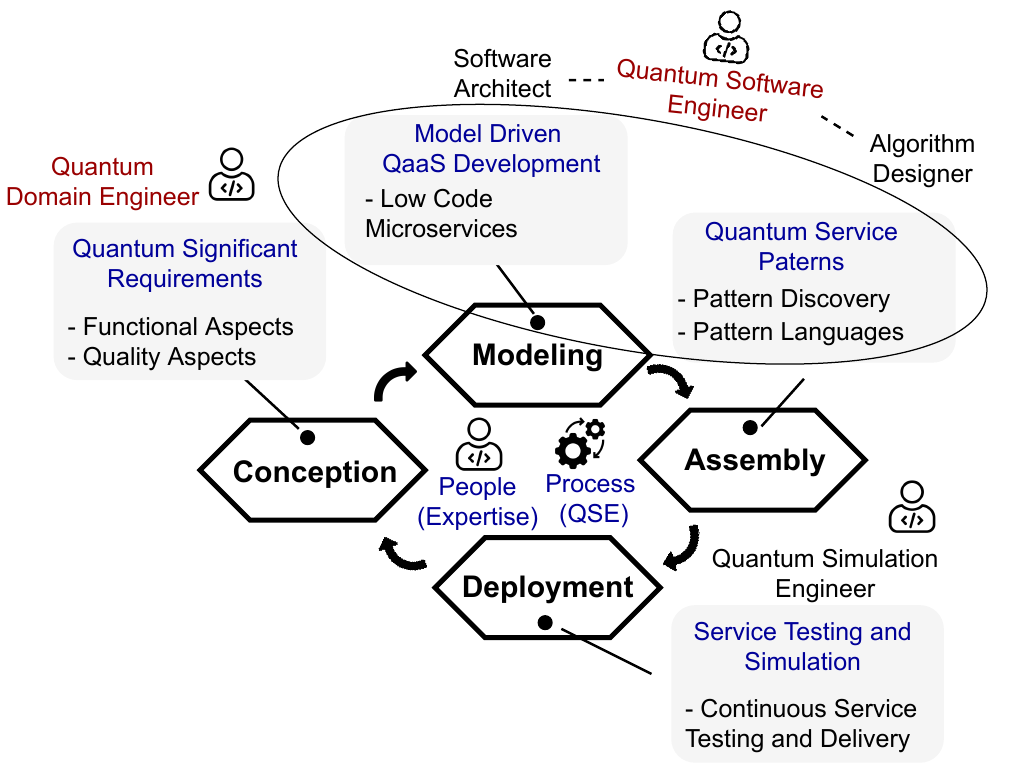} 
 	\caption{Overview of Emerging Research Trends}
	\label{FutureResearch}
\end{figure}

We now answer RQ-2 that aims to report the emerging trends that indicate dimensions of potential future research on QCaaS. To identify these trends, we specifically reviewed the details about objectives/contributions, evaluation/demonstrations, and limitations/future research in each study during quality assessment (Q-1, Q-4, Q-5 in Table \ref{tab:qualitycriteria}). We have visualised the identified emerging trends in Figure \ref{FutureResearch} as per the SOA life cycle activities. For example, Figure \ref{FutureResearch} highlights that during the conception of quantum services, Quantum Significant Requirements (QSRs) should include quality aspects that complement the functional aspects to ensure the required functionality and desired quality of QCaaS as part of quantum domain engineering activity in QSE. 



\subsection{Process and Human Roles in QCaaS Development} 
Process-centred engineering is manifested in service life cycle to structure a multitude of activities, such as service design, development, and delivery, into a coherent process that can be executed in an incremental fashion \cite{R19_ahmadtowards}. Although the evidence from the selected studies is accumulated and organized under SOA life cycle in Table \ref{tab:criteria}, however, at the individual scale, the studies lacked a process-centric approach and highlighted the needs for processes attuned to the needs of QSE \cite{R19_ahmadtowards}. These processes, such as Quantum DevOps, Quantum  microservicing or agile methods for quantum service/software etc., are seen as quantum-genre of SE processes that are better aligned with the needs of QSE \cite{R23_khan2022agile}. For example, in the agile method for QSE, quantum domain engineering can help accumulate the domain information of a quantum system (hardware/software) to develop a design model that can act as a blueprint to implement quantum software and services. Process-centric approaches can also support tools (for automation) and professional roles (human decision support) to engineer quantum services. We identified the need for human roles as QSE professionals to effectively undertake QCaaS development.

\faComment{} 
{~\footnotesize\textsf{{\textbf{Process-centred} approaches, such as Quantum DevOps, Quantum microservicing or agile methods, can enable an incremental development and delivery of quantum software services. \textbf{Human roles} can enrich the process - synergising knowledge of quantum physics practices of software engineering - to support activities of quantum domain engineering, software architecting, and simulation management as expertise that currently lacks in quantum service-orientation.}}}

\subsection{Quantum Significant Requirements}   
The concept of Quantum Significant Requirements (QSRs) can be traced to well-established role of Architecturally Significant Requirements (ASRs) in software and service engineering. QSRs as part of the quantum domain engineering activity can help elicit and document the functional and non-functional aspects (also quality aspects or attributes) of quantum software. Existing research primarily focuses on functional aspects of services while overlooking the non-functional aspects that include but are not limited to QuBit utilisation, energy efficiency, quantum vendor lock-ins, QPU elasticity, quantum error mitigation, that impact the development and operationalisation or QCaaS solutions. For example, during the quantum domain engineering activity the hardware aspects (operations of QuGates) can be mapped to software aspects (components and connectors) that can help with splitting the computation tasks between classical and quantum computers using the classic-quantum split pattern. 

\faComment{} 
{~\footnotesize\textsf{The notion of {\textbf{QSRs} is rooted in the concept of ASRs that aims to identify, classify, and document functional and quality aspects of quantum software services. QSRs for quantum services can complement the functional aspects with quality requirements to ensure the required functionality and desired quality of service.}}} 

\subsection{Model-driven Quantum Software Servicing}
Model-driven Service Engineering (MDSE) enables software engineers and architects to rely on models that can abstract complex and implementation-specific details with human comprehensible visual notations to design and implement software services. Specifically, by exploiting MDSE, software engineers can apply the model transformation to transform the design models into implementation (source code) and validation (test case) models. Modeling notations, such as Service-oriented architecture Modeling Language (SoaML) and more specifically the Q-UML specification, provide a metamodel and a UML profile for the specification and design of services within a service-oriented architecture. MDSE can benefit novice developers to to map the flow of algorithms and modules of source code to graphical models for low-code (model-driven) development of quantum services.

\faComment{}
{~\footnotesize\textsf{\textbf{Model-driven Quantum Service Development} can leverage the principle of MDSE and modeling notations like SoaML and Q-UML to abstract implementation-level complexities with design-level models driven by QSRs. To exploit model-driven development, there is a need for tool support that could enable automation (e.g., model transformation tools) and human decision support (e.g., quantum software architects) to quantum software servicing using MDSE.}}

\begin{table*}[hbt!]
\caption{A Review of the Relevant Secondary Studies}
\begin{centering}
{\tiny
\begin{tabular}{|lcclcl|}
\hline
\rowcolor[HTML]{DAE8FC} 
\multicolumn{6}{|c|}{\cellcolor[HTML]{DAE8FC}\textbf{Quantum Software Engineering}} \\ \hline
\rowcolor[HTML]{EFEFEF} 
\multicolumn{1}{|l|}{\cellcolor[HTML]{EFEFEF}\textit{\begin{tabular}[c]{@{}l@{}}Study \\ Reference\end{tabular}}} & \multicolumn{1}{c|}{\cellcolor[HTML]{EFEFEF}\textit{\begin{tabular}[c]{@{}c@{}}Type of \\ Study  \end{tabular}}} & \multicolumn{1}{c|}{\cellcolor[HTML]{EFEFEF}\textit{\begin{tabular}[c]{@{}c@{}}Focus of \\ Study\end{tabular}}} & \multicolumn{1}{c|}{\cellcolor[HTML]{EFEFEF}\textit{\begin{tabular}[c]{@{}c@{}}Core\\ Findings\end{tabular}}} & \multicolumn{1}{l|}{\cellcolor[HTML]{EFEFEF}\begin{tabular}[c]{@{}l@{}}QSE\\ Activity\end{tabular}} & \begin{tabular}[c]{@{}l@{}}Publication \\ Year\end{tabular} \\ \hline
\multicolumn{1}{|l|}{\cite{X1_QSE}} & \multicolumn{1}{c|}{\begin{tabular}[c]{@{}c@{}}Literature \\ Review\end{tabular}} & \multicolumn{1}{c|}{\begin{tabular}[c]{@{}c@{}}Quantum \\ Software \\ Engineering\end{tabular}} & \multicolumn{1}{l|}{\begin{tabular}[c]{@{}l@{}}QSE life cycle, quantum software \\ engineering processes, methods, and tools.\end{tabular}} & \multicolumn{1}{c|}{\begin{tabular}[c]{@{}c@{}}Software  \\ Life cycle\end{tabular}} & 2020 \\ \hline
\multicolumn{1}{|l|}{\cite{X2_QSA}} & \multicolumn{1}{c|}{\begin{tabular}[c]{@{}c@{}}Systematic \\ Review\end{tabular}} & \multicolumn{1}{c|}{\begin{tabular}[c]{@{}c@{}}Quantum\\ Software \\ Architecture\end{tabular}} & \multicolumn{1}{l|}{\begin{tabular}[c]{@{}l@{}}Quantum software architecting activities, \\ modelling notations, patterns, and tools \\ for architectural development\end{tabular}} & \multicolumn{1}{c|}{\begin{tabular}[c]{@{}c@{}}Software Design \\ and \\ Architecture\end{tabular}} & 2021 \\ \hline
\multicolumn{1}{|l|}{\cite{R15_de2022software}} & \multicolumn{1}{c|}{\begin{tabular}[c]{@{}c@{}}Repository Mining\\ and   \\ Practitioner Survey\end{tabular}} & \multicolumn{1}{c|}{\begin{tabular}[c]{@{}c@{}}Quantum\\ Programming\\ Languages\end{tabular}} & \multicolumn{1}{l|}{\begin{tabular}[c]{@{}l@{}}Mining repositories and interviewing \\ practitioner to investigate quantum \\ programming languages\end{tabular}} & \multicolumn{1}{c|}{\begin{tabular}[c]{@{}c@{}}Software \\ Implementation\end{tabular}} & 2022 \\ \hline
\multicolumn{1}{|l|}{\cite{x4_Issues}} & \multicolumn{1}{c|}{\begin{tabular}[c]{@{}c@{}}Repository \\ Mining\end{tabular}} & \multicolumn{1}{c|}{\begin{tabular}[c]{@{}c@{}}Quantum\\ Issues\end{tabular}} & \multicolumn{1}{l|}{\begin{tabular}[c]{@{}l@{}}Mining repositories to identify technical\\ debts in open-source quantum software\end{tabular}} & \multicolumn{1}{c|}{\begin{tabular}[c]{@{}c@{}}Software \\ Implementation \\ and   Testing\end{tabular}} & 2022 \\ \hline
\multicolumn{1}{|l|}{\cite{X5_QC}} & \multicolumn{1}{c|}{\begin{tabular}[c]{@{}c@{}}Vision \\  Paper\end{tabular}} & \multicolumn{1}{c|}{\begin{tabular}[c]{@{}c@{}}Quantum\\ Software \\ Architecture\end{tabular}} & \multicolumn{1}{l|}{\begin{tabular}[c]{@{}l@{}}QSE life cycle and comparing   quantum\\ computers with their classical \\ counterparts and vision for future research\end{tabular}} & \multicolumn{1}{c|}{\begin{tabular}[c]{@{}c@{}}Software  \\ Life cycle\end{tabular}} & 2022 \\ \hline
\rowcolor[HTML]{DAE8FC} 
\multicolumn{6}{|c|}{\cellcolor[HTML]{DAE8FC}\textbf{Quantum Services Computing}} \\ \hline
\multicolumn{1}{|l|}{\cite{R11_leymann2020quantum}} & \multicolumn{1}{c|}{\begin{tabular}[c]{@{}c@{}}Literature \\ Survey\end{tabular}} & \multicolumn{1}{c|}{\begin{tabular}[c]{@{}c@{}}Quantum \\ Cloud \\ Computing\end{tabular}} & \multicolumn{1}{l|}{\begin{tabular}[c]{@{}l@{}}Programming quantum computers and \\ investigating hybrid software consisting \\ of classical parts  and quantum parts.\end{tabular}} & \multicolumn{1}{c|}{\begin{tabular}[c]{@{}c@{}}Service Design \\ and \\ Architecture\end{tabular}} & 2020 \\ \hline
\multicolumn{1}{|l|}{\cite{x7_QAAS}} & \multicolumn{1}{c|}{\begin{tabular}[c]{@{}c@{}}Literature\\ Survey\end{tabular}} & \multicolumn{1}{c|}{\begin{tabular}[c]{@{}c@{}}Quantum \\ Service \\ Computing\end{tabular}} & \multicolumn{1}{l|}{\begin{tabular}[c]{@{}l@{}}Potential and limitation  of integrating \\ quantum computing with cloud computing\end{tabular}} & \multicolumn{1}{c|}{\begin{tabular}[c]{@{}c@{}}Service \\ Lifecycle\end{tabular}} & 2015 \\ \hline
\end{tabular}}
\par\end{centering}
\label{tab:RelatedWork}
\end{table*}
\subsection{Empiricism in Mining Quantum Service Patterns}
Pattern-based software service engineering relies on architectural design and implementation strategies and best practices that can be reused to deliver software services \cite{R20_keen2006patterns}. Existing solutions employ a number of pattern-based solutions, such as classic-quantum split and service wrapping pattern, however, there is no evidence on an empirical discovery of patterns and tactics as reusable knowledge \cite{R27_GitQSE}. A lack of empiricism in discovering new patterns hinders the reusability of service design and implementation knowledge during quantum service engineering. One possible dimension for pattern discovery is mining software repositories or social coding platforms (e.g., GitHub) that contain raw knowledge that can be mined  as patterns. Pattern-based solutions could complement human expertise with available best practices for service design and implementation. 

\faComment{}
{~\footnotesize\textsf{\textbf{Pattern discovery} via empirically-grounded methods, i.e., mining repositories or social coding platforms of quantum software development, can help leverage reusable design rationale for quantum service engineering. Pattern languages can empower the role of quantum software architects and algorithm designers to rely on reusable knowledge and best practices as opposed to ad-hoc and once-off solutions.}}


\subsection{Continuous Testing and Delivery of Quantum Services} 
With an adoption of agile software engineering in quantum software development context \cite{R23_khan2022agile}, there is a need for light and adaptive methods to ensure a continuous development and delivery of quantum software services. The literature suggested a lack of solutions on testing the software services. The continuous testing and continuous delivery can help CT/CD to test the services against QSRs more effectively and deliver them rapidly. Quantum service testing can involve simulation or regression tests against the QSRs.


\faComment {~\footnotesize\textsf{\textbf{Continuous Testing and Continuous Delivery} (CT/CD) relies on agile software engineering methods to help deliver quantum software services rapidly and reliably. CT/CD can provide strategic benefits to vendors by adding new services to their quantum platforms.}}


\section{Related Survey-based and Empirical Studies}

We now review the relevant research rooted in survey-based evidence or empirical studies on (i) engineering and architecting quantum software (Section \ref{Related:QSE}), and (ii) quantum services computing (Section \ref{Related:QAAS}), as in Table \ref{tab:RelatedWork}. 

\textbf{Interpretation of the Review}: To compare and summarise objectively, Table \ref{tab:RelatedWork} highlights each study using five-point self-explanatory criteria including (a) type of study (adopted from \cite{R16_petersen2008systematic}), (b) focus of study, (c) core findings, (d) QSE activity supported by the study, and (e) year of publication, each exemplified below. For example, the study \cite{X2_QSA} presents a systematic review, available since 2021, as part of evidence-based software engineering to focus on architecting quantum software. It presents the core findings about architectural life cycle and state-of-the-art on architectural modeling, patterns, and tools to architects develop quantum software. 

\subsection{Architecting and Engineering Quantum Software}\label{Related:QSE}
Quantum Software Engineering (QSE) has emerged as the most recent genre of software engineering (SE) that allows practitioners to adopt a process-centric and systematic approach develop software-intensive systems and applications for QC \cite{X1_QSE, X5_QC}. Some recently published survey-based studies on QSE highlight that existing SE principles and practices can be tailored to develop quantum software, however, issues specific to QC, such as operationalising QuBits/QuGates, quantum domain engineering, and quantum-classic software split, require new engineering methods to tackle these challenges \cite{X5_QC}. To derive new methods and processes, the QSE research community is striving to organize a body of knowledge and academic collaborations that can be stimulated via dedicated publication forums. Academic discussions at the publication fora and published results \cite{R1_ali2022software}\cite{R4_dyakonov2019will} have highlighted that in addition to the engineering principle and practices, human knowledge and expertise require fundamental knowledge of quantum mechanics to effectively design algorithmic solutions for QSE projects \cite{R15_de2022software}. The current generation of software architects and developers who lack the foundational knowledge of quantum mechanics may be hindered or find themselves under-prepared for quantum software development \cite{X2_QSA}. Software architecture is being viewed as a solution that can abstract complexities of implementation by representing software to be developed as architectural components and connectors. A recently conducted systematic review of architectural solutions puts forward five architecting activities to guide software designers to engineer quantum software solutions in an incremental manner \cite{X2_QSA}. 

\subsection{Quantum Services Computing}\label{Related:QAAS}
Quantum service-orientation is a recent trend initially pushed by QC vendors to allow developers who can compose and invoke software services on remotely hosted quantum systems and infrastructures \cite{R11_leymann2020quantum}. Quantum service computing initiatives, such as Amazon Braket and Q experience, have paved the way for academic research to propose solutions for quantum cloud, quantum services computing, and quantum as a service. Recent research reviews inform about the current progress and emerging challenges for quantum services computing that include but are not limited to hardware availability, quantum noise, and quantum-classic split \cite{x7_QAAS}. Software researchers are striving to synergise existing methods of microservicing and quantum software development such as Quantum DevOps to develop quantum microservices \cite{R23_khan2022agile}. 

\section{Conclusions and Future Work}\label{conclusions}
Quantum service orientation allows QC vendors to offer end-users’ access to computational resources - enabling shots at QPUs - by following the utility computing model. Recently, research and development on QSE have started to synergize the principles of service-orientation and practices of QC (algorithms, simulations, QuGates etc.) to enable the adoption of QCaaS by individuals and enterprises to attain quantum supremacy in modern day computing. 
To this end, we used the systematic mapping study approach to investigate (i) existing solutions for quantum service development that enable or enhance QCaaS, and (ii) emerging trends that highlight the needs for ongoing and future research on QCaaS. The results of this SMS as structured information (Table \ref{tab:criteria}) and visual illustrations (Figure \ref{Fig-1:Context}-\ref{Results}) highlight the strengths, limitations, and potential for future research from the QSE perspective.  

\textbf{Implications and Future Work}: The primary implication of this SMS is to establish an evidence-based body of knowledge for service development that leverages design notations, patterns, and architectural approaches highlighting the needs for human-centric and process-driven QCaaS development. The results of this SMS establish the foundations for future work in three directions that include (a) \textit{conducting practitioners' survey}, (b) \textit{implementing the reference architecture}, and (c) \textit{mining social coding platforms} for empiricism in QCaaS computing. The literary foundations can help us to design an empirical study that aims to engage service developers and engineers in a workshop (activity and survey) to seek their feedback and synthesise the results as practitioners' perspectives to complement the evidence from academic research. We also plan to mine social coding platforms (e.g., GitHub) to empirically discover knowledge and understand the practices adopted by developers' communities in open-source QCaaS.

{\renewcommand{\arraystretch}{1}
\begin{table}[h]
\centering
\tiny
\caption{List of the selected studies of this SMS}
\label{tab:SelectedStudies}
\begin{tabular}{|c|l|c|c|}
\hline
\rowcolor[HTML]{DAE8FC} 
\textbf{\begin{tabular}[c]{@{}c@{}}Study\\ ID\end{tabular}} &
  \multicolumn{1}{c|}{\cellcolor[HTML]{DAE8FC}\textbf{\begin{tabular}[c]{@{}c@{}}Authors, Title \\ and Venue\end{tabular}}} &
  \textbf{\begin{tabular}[c]{@{}c@{}}Publication\\ Year\end{tabular}} &
  \textbf{\begin{tabular}[c]{@{}c@{}}Quality\\ Score\end{tabular}} \\ \hline
{[}S1{]} &
  \begin{tabular}[c]{@{}l@{}}G. Jose, J. Rojo, D. Valencia, E. Moguel, J. Berrocal,\\    and Juan. Murillo. \textsf{Quantum Software as a Service}\\  \textsf{through a Quantum API Gateway. \textit{IEEE Internet Computing}}\end{tabular} &
  2021 &
  4.0 \\ \hline
{[}S2{]} &
  \begin{tabular}[c]{@{}l@{}}Kumara, Indika, Willem-Jan Van Den Heuvel, and \\ Damian A. Tamburri. QSOC: \textsf{Quantum  service}\\ \textsf{oriented computing. \textit{SummerSOC}}\end{tabular} &
  2021 &
  3.0 \\ \hline
{[}S3{]} &
  \begin{tabular}[c]{@{}l@{}}Nguyen, Hoa T., Muhammad Usman, and Rajkumar \\ Buyya. \textsf{QFaaS: A Serverless Function-as-a-Service} \\ \textsf{Framework for Quantum Computing. \textit{arXiv}}\end{tabular} &
  2022 &
  5.0 \\ \hline
{[}S4{]} &
  \begin{tabular}[c]{@{}l@{}}Moguel, Enrique, Javier Rojo, David Valencia, Javier \\ Berrocal, Jose Garcia-Alonso, and Juan M. Murillo.\\ \textsf{Quantum service-oriented computing: current landscape}\\  \textsf{and challenges. \textit{Software Quality Journal}}\end{tabular} &
  2022 &
  4.0 \\ \hline
{[}S5{]} &
  \begin{tabular}[c]{@{}l@{}}Rojo, Javier, David Valencia, Javier Berrocal, Enrique \\ Moguel, Jose Garcia-Alonso, and Juan Manuel Murillo \\ Rodriguez. \textsf{Trials and tribulations of developing hybrid}\\ \textsf{quantum-classical microservices systems. \textit{Q-SET}}\end{tabular} &
  2021 &
  3.5 \\ \hline
{[}S6{]} &
  \begin{tabular}[c]{@{}l@{}}De Stefano, Manuel, Dario Di Nucci, Fabio Palomba, \\ Davide Taibi, and Andrea De Lucia. \textsf{Towards Quantum}\\ \textsf{algorithms-as-a-service. \textit{QP4SE}}\end{tabular} &
  2022 &
  2.5 \\ \hline
{[}S7{]} &
  \begin{tabular}[c]{@{}l@{}}Valencia, David, Enrique Moguel, Javier Rojo, Javier \\ Berrocal, Jose Garcia-Alonso, and Juan M. Murillo. \\ \textsf{Quantum Service-Oriented Architectures: From Hybrid}\\ \textsf{Classical Approaches to Future Stand-Alone Solutions.}\\ \textit{Quantum Software Engineering}\end{tabular} &
  2022 &
  3.5 \\ \hline
{[}S8{]} &
  \begin{tabular}[c]{@{}l@{}}Barzen, Johanna, Frank Leymann, Michael Falkenthal, \\ Daniel Vietz, Benjamin Weder, and Karoline Wild. \\ \textsf{Relevance of near-term quantum computing in the} \\ \textsf{cloud: A humanities perspective. \textit{CLOSER}}\end{tabular} &
  2021 &
  4.0 \\ \hline
{[}S9{]} &
  \begin{tabular}[c]{@{}l@{}}Valencia, David, Jose Garcia-Alonso, Javier Rojo, \\ Enrique Moguel, Javier Berrocal, and Juan \\ Manuel Murillo. \textsf{Hybrid classical-quantum} \\ \textsf{software services systems: Exploration of the}\\ \textsf{rough edges.} \textit{Quality of Information and} \\ \textit{Communications Technology}\end{tabular} &
  2021 &
  3.5 \\ \hline
\end{tabular}
\end{table}}

\balance
\bibliographystyle{ieeetr}
\bibliography{References}


\end{document}